\documentclass[a4,11pt,dvips]{article}

\usepackage{amsmath,epsfig,ulem}
\usepackage{url}
\usepackage{rotating}
\usepackage{cite}
\usepackage{hhline}
\usepackage{array}
\usepackage{amssymb}
\usepackage{color}
\usepackage{dsfont}
\usepackage{axodraw}

\begin{document}
\tolerance=1000000

\newcommand{\imag}{\Im {\rm m}}
\newcommand{\real}{\Re {\rm e}}

\def\tablename{\bf Table}%
\def\figurename{\bf Figure}%

\newcommand{\sts}{\scriptstyle}
\newcommand{\ngs}{\!\!\!\!\!\!}
\newcommand{\rb}[2]{\raisebox{#1}[-#1]{#2}}
\newcommand{\CP}{${\cal CP}$~}
\newcommand{\sbomu}{\frac{\sin 2 \beta}{2 \mu}}
\newcommand{\kmol}{\frac{\kappa \mu}{\lambda}}
\newcommand{\s}{\\ \vspace*{-3.5mm}}
\newcommand{\lsim}{\raisebox{-0.13cm}{~\shortstack{$<$\\[-0.07cm] $\sim$}}~}
\newcommand{\gsim}{\raisebox{-0.13cm}{~\shortstack{$>$\\[-0.07cm] $\sim$}}~}
\newcommand{\kr}{\color{red}}

\newcommand{\hdick}{\noalign{\hrule height1.4pt}}
\newcommand{\eV}  {\mathrm{eV}}
\newcommand{\keV} {\mathrm{keV}}
\newcommand{\MeV} {\mathrm{MeV}}
\newcommand{\GeV} {\mathrm{GeV}}
\newcommand{\TeV} {\mathrm{TeV}}
\newcommand{\fb}  {\mathrm{fb}}
\newcommand{\fbi} {\mathrm{fb}^{-1}}
\newcommand{\ab}  {\mathrm{ab}}
\newcommand{\abi} {\mathrm{ab}^{-1}}
\newcommand{\mrad}{\mathrm{mrad}}
\newcommand{\cm}  {\mathrm{cm}}
\newcommand{\sek} {\mathrm{s}}
\newcommand{\msek}{\mathrm{ms}}
\newcommand{\cO}  {{\cal O}}
\newcommand{\cL } {{\cal L}}
\newcommand{\cP } {{\cal P}}
\newcommand{\cB } {{\cal B}}
\newcommand{\cF } {{\cal F}}

\def\ee    {e^+e^-}
\def\ti    {\tilde}
\def\sf    {{\ti f}}
\def\sq    {{\ti q}}
\def\stau  {{\ti\tau}}
\def\sell  {{\ti\ell}}
\def\cx    {{\ti\chi}}
\def\ch    {{\ti\chi}}
\def\nt    {{\ti\chi^0}}
\def\smu   {{\ti\mu}}
\def\smul  {{\ti\mu_L}}
\def\smur  {{\ti\mu_R}}
\def\snm   {{\ti\nu_\mu}}
\def\se    {{\ti e}}
\def\sel   {{\ti e_L}}
\def\ser   {{\ti e_R}}
\def\sne   {{\ti\nu}_e}
\def\snl   {{\ti\nu}_l}
\def\snu   {{\ti\nu}}
\def\snt   {{\ti\nu}_\tau}

\def \Eslash {E \kern-.6em\slash }
\newcommand{\eq}[1]{eq.~(\ref{#1})}
\newcommand{\fig}[1]{Fig.~\ref{#1}}
\newcommand{\tab}[1]{Table~\ref{#1}}


\renewcommand{\theequation}{{\rm \thesection.\arabic{equation}}}

\begin{titlepage}

\begin{flushright}
DESY 06-239\\
KEK-TH-1126\\
KIAS-P06064\\
\today
\end{flushright}

\vskip 1.5cm

\begin{center}
{\large \bf SPIN ANALYSIS OF SUPERSYMMETRIC PARTICLES}\\[0.7cm]
{\normalsize S.Y. Choi$^{1,2}$, K. Hagiwara$^{3}$, H.-U. Martyn$^{1,4}$,
        K. Mawatari$^{5}$ and P.M. Zerwas$^{1,3}$}\\[0.8cm]
{\it $^1$ Deutsches Elektronen-Synchrotron DESY, D-22603 Hamburg, Germany\\
     $^2$ Physics Department and RIPC, Chonbuk University, Jeonju 561-756,
          Korea\\
     $^3$ Theory Division, KEK, Tsukuba, Ibaraki 305-0801, Japan\\
     $^4$ I. Physikalisches Institut, RWTH Aachen, Aachen, Germany\\
     $^5$ School of Physics, Korea Institute for Advanced Study,
          Seoul 130-722, Korea}
\end{center}

\renewcommand{\thefootnote}{\arabic{footnote}}
\vspace{1.8cm}

\begin{abstract}
\noindent The spin of supersymmetric particles can be determined at
$e^+e^-$ colliders unambiguously. This is demonstrated for a
characteristic set of non-colored supersymmetric particles -- smuons,
selectrons, and charginos/neutralinos. The analysis is based on the
threshold behavior of the excitation curves for pair production in $e^+e^-$
collisions, the angular distribution in the production process and
decay angular distributions. In the first step we present the
observables in the helicity formalism for the supersymmetric
particles. Subsequently we confront the results with corresponding
analyses of Kaluza-Klein particles in theories of universal extra
space dimensions which behave distinctly different from
supersymmetric theories. It is shown in the third step that a set of
observables can be designed which signal the spin of supersymmetric
particles unambiguously without any model assumptions. Finally in
the fourth step it is demonstrated that the determination of the
spin of supersymmetric particles can be performed experimentally in
practice at an $e^+e^-$ collider.
\end{abstract}

\end{titlepage}

\newpage

\section{INTRODUCTION}
\label{sec:introduction}
\setcounter{equation}{0}

The spin is one of the characteristics of all particles and it must be
determined experimentally for any new species. Compelling arguments have
been forwarded which suggest the supersymmetric extension of the Standard
Model \cite{Wess:1974tw,nilles_haber_kane}. In supersymmetric theories (SUSY)
spin-1 gauge and spin-0 Higgs bosons are paired with
spin-1/2 fermions, gauginos and higgsinos, which mix, in the
non-colored sector, to form charginos and
neutralinos. Analogously spin-1/2 leptons and quarks are paired with
spin-0 scalar sleptons and squarks. This opens a wide area of
necessary efforts to determine the nature of the new particles
experimentally.\s

Measuring the masses of the particles is not sufficient to unravel the
nature of the particles and of the underlying theory. This point has been
widely discussed by comparing supersymmetric theories with theories
of universal extra space dimensions (UED) \cite{appelquist,Cheng:2002iz}
in which the
counterparts of the supersymmetric partners are Kaluza-Klein (KK)
excitations of the standard particles. When supersymmetric squarks are
produced \cite{beenakker} at LHC, they may cascade down \cite{LHCILC} to
standard particles in the chain
$\tilde{q}\to q\tilde{\chi}^0_2\to q\bar{\ell}\tilde{\ell}\to
q\bar{\ell}\ell\tilde{\chi}^0_1$, which generates the observable final state
$q\bar{\ell}\ell$. However, an analogous cascade can be realized in
theories of universal extra space dimensions, starting from
a KK excitation $q_1$ of a quark,
$q_1\to qZ_1\to q\bar{\ell}\ell_1\to q \bar{\ell}\ell\gamma_1$ \cite{datta}.
The origin of the observed chain particles, supersymmetry or extra space
dimensions, can clearly be unraveled by measuring the spins of
the intermediate cascade particles.\s

Spin measurements of supersymmetric particles are difficult at
LHC \cite{datta,barr_webber,LHCtau}. While the invariant mass distributions of
the particles in decay cascades are characteristic for the spins
of the intermediate particles involved, detector effects
strongly reduce the signal in practice.\s

In contrast, several techniques can be exploited to determine
unambiguously the spin of particles produced pairwise in $e^+e^-$ collisions.
These techniques have first been worked out
theoretically for Higgs bosons, studied in the Higgs-strahlung
process \cite{choi_eberle}; subsequent experimental simulations
have proven these techniques to work in practice \cite{lohmann}.
To conform with its scalar character,
the polar angle distribution in smuon pair production has been
investigated directly by reconstruction in Ref.~\cite{martyn}
and reflected in their decay products in Ref.~\cite{battaglia} at TeV
and multi-TeV $e^+e^-$ colliders, respectively.\s

A sequence of techniques, increasing in complexity, can be applied
to determine the spin of particles in pair production
\begin{eqnarray}
  e^+e^-\ \ \to\ \ \tilde{\mu}^+\tilde{\mu}^- , \, \tilde{e}^+ \tilde{e}^- \;\;
  {\rm and} \;\;\,
          \tilde{\chi}^+\tilde{\chi}^-  , \,
          \tilde{\chi}^0\tilde{\chi}^0
\end{eqnarray}
of sleptons, charginos and neutralinos in $e^+e^-$ collisions:\\[-0.7cm]
\begin{itemize}
\item[{ }] (a) rise of the excitation curve near the threshold;\\[-0.8cm]
\item[{ }] (b) angular distribution in the production process;\\[-0.8cm]
\item[{ }] (c) angular distribution in decays of the polarized
           particles,
\end{itemize}
eventually supplemented by
\begin{itemize}
\item[{ }] (d) angular correlations between decay products of two
           particles.\\[-0.7cm]
\end{itemize}
While the second step (b) is already sufficient in the slepton sector, only
the final state analysis is sufficient in general, including charginos/neutralinos,
to determine the spin unambiguously. On the experimental side we follow the
standard path. It will be shown in detailed
simulations that the theoretically predicted distributions in
supersymmetric theories can be reconstructed after including initial and
final state QED radiation,
beamstrahlung and detector effects. Within the extended theoretical
frame it is then proven that the assignment of the spin is
unambiguous indeed.\s

The report is organized as follows. In the subsequent Sections 2 to 4
we set up the theoretical basis for spin measurements of smuons, selectrons,
and charginos/neutralinos. The technical frame we have chosen is the
helicity formalism. We analyze which observables must be
measured to determine the spin unambiguously. Moreover, simulations
will assure us that the analyses of supersymmetric theories in
$e^+e^-$ collisions can be performed experimentally. In the last
Section 5 we briefly summarize the results. General formulae for the production
cross sections of supersymmetric particles in collisions of polarized
electrons and positrons are presented in an Appendix.\s

\section{SPIN OF SMUONS}
\label{sec:smuon_spin}
\setcounter{equation}{0}

\subsection{Smuon Production in \boldmath{$e^+e^-$} Collisions}

\begin{figure}[htb]
{
\begin{center}
\begin{picture}(250,80)(30,20)
\Text(0,80)[r]{\bf (a)}
\Text(25,75)[r]{$e^-$}
\ArrowLine(30,75)(55,50)
\ArrowLine(55,50)(30,25)
\Text(25,25)[r]{$e^+$}
\Photon(55,50)(97,50){2}{10}
\Text(75,37)[]{\color{blue} $\gamma,\,Z$}
\Line(95,50)(120,75)
\Line(97,50)(121,74)
\Text(135,75)[]{$\tilde{\ell}^-$}
\Line(120,25)(95,50)
\Line(121,26)(97,50)
\Text(135,25)[]{$\tilde{\ell}^+$}
\Text(185,80)[r]{\bf (b)}
\Text(200,75)[]{$e^-$}
\ArrowLine(210,75)(255,75)
\Text(200,25)[]{$e^+$}
\ArrowLine(255,25)(210,25)
\Line(254,75)(254,25)
\Photon(254,75)(254,25){2}{9}
\Text(244,50)[]{\color{blue} $\tilde{\chi}^0$}
\Line(255,75)(300,75)
\Line(255,73.5)(300,73.5)
\Text(310,75)[l]{$\tilde{e}^-$}
\Line(300,25)(255,25)
\Line(300,26.5)(255,26.5)
\Text(310,25)[l]{$\tilde{e}^+$}
\end{picture}
\end{center}
}
\caption{\it (a) $s$-channel $\gamma$ and $Z$ exchange diagrams contributing
  to the production of all slepton pairs in $e^+e^-$ annihilation; and (b)
  $t$-channel neutralino exchange diagram contributing only to
  the production of selectron pairs in $e^+e^-$ collisions.}
\label{fig:diagram1}
\end{figure}
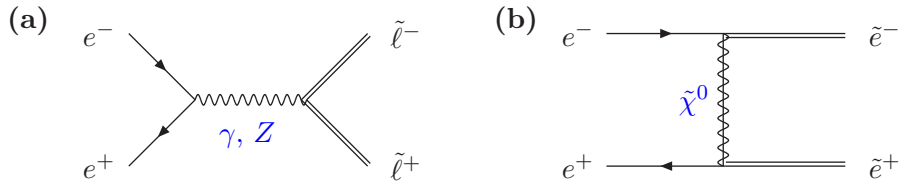

Smuons are the prototype for scalar particle pair production in $e^+e^-$
collisions \cite{ILC,CLIC}
mediated by the $s$-channel exchange of $\gamma$ and $Z$ boson. Different
lepton numbers prevent the flow of particles from the initial to the
final state. For the sake of [experimental] simplicity we will
restrict ourselves to the analysis of R-type smuons,
\begin{eqnarray}
  e^+e^-\ \ \rightarrow\ \ \tilde{\mu}^+_R\, \tilde{\mu}^-_R
\end{eqnarray}
as these particles almost exclusively decay through the 2-particle channel
$\tilde{\mu}^\pm_R\to \mu^\pm \tilde{\chi}^0_1$ with only one escaping invisible
particle. The process is described by diagram (a) in Fig.$\,$\ref{fig:diagram1}.\s

The amplitude describing this production process can be expressed
in terms of the generalized electron charges
\begin{eqnarray}
&& Q_L = 1+\left(s^2_W-1/2\right) c^{-2}_W D_Z(s) \label{eq:smuon_Q_L}\\
&& Q_R = 1+ t^2_W D_Z(s) \label{eq:smuon_Q_R}
\end{eqnarray}
with $s^2_W =\sin^2\theta_W$ {\it etc}, $\theta_W$ being the electroweak
mixing angle, and the normalized $Z$ propagator $D_Z(s)=s/[s-m^2_Z+i m_Z\Gamma_Z]$,
$s$ denoting the squared center-of-mass  energy [$D_Z$ is approximately
real in the high energy limit $s\gg m^2_Z$]. The indices $L$ and $R$ in
Eqs.$\,$(\ref{eq:smuon_Q_L}) and (\ref{eq:smuon_Q_R}) refer to left- and
right-handedly polarized electrons [and oppositely polarized positrons],
respectively.\s

The total cross section and the distribution in the polar angle $\theta$
between the $\tilde{\mu}^\pm$ flight direction  and the $e^+e^-$ beam axis can
be written in the form\footnote{The complete set of 1-loop radiative
corrections, including the genuine SUSY corrections has been presented
in Refs.~\cite{RC1,RC2}; see also Ref.~\cite{SPA}.}
\begin{eqnarray}
\sigma\left[e^+e^-\to\tilde{\mu}^+_R\tilde{\mu}^-_R\right] &=&
   \frac{\pi\alpha^2}{6s} \beta^3 \left[Q^2_L + Q^2_R\right]
  \label{eq:smur_total_cross_section}
   \\[2mm]
\frac{1}{\sigma}\frac{d\sigma}{d\cos\theta}
\left[e^+e^-\to\tilde{\mu}^+_R\tilde{\mu}^-_R\right] &=&
  \frac{3}{4} \sin^2\theta
  \label{eq:smur_differential_cross_section}
\end{eqnarray}
The coefficient $\beta^3$, with $\beta=(1-4 m^2_{\tilde{\mu}_R}/s)^{1/2}$
denoting the velocity of the smuons, is the product of the phase space suppression
factor $\beta$, and the square of the $P$-wave suppression $\sim\beta$ near
the threshold. The scalar smuon pair is produced in a $P$-wave to balance
the spin 1 of the intermediate vector boson. Angular momentum conservation
leads also to the $\sin^2\theta$ dependence of the differential cross
section as forward production of spinless particles is forbidden.\s

For asymptotic energies the cross section
\begin{equation}
\sigma\ \ \to\ \  \frac{5 \pi \alpha^2}{24 c_W^4}\,\frac{1}{s} \quad
   \mbox{for}\quad  s\ \ \rightarrow\ \ \infty
\end{equation}
follows the appropriate scaling law. \s

\begin{figure}[ht!]
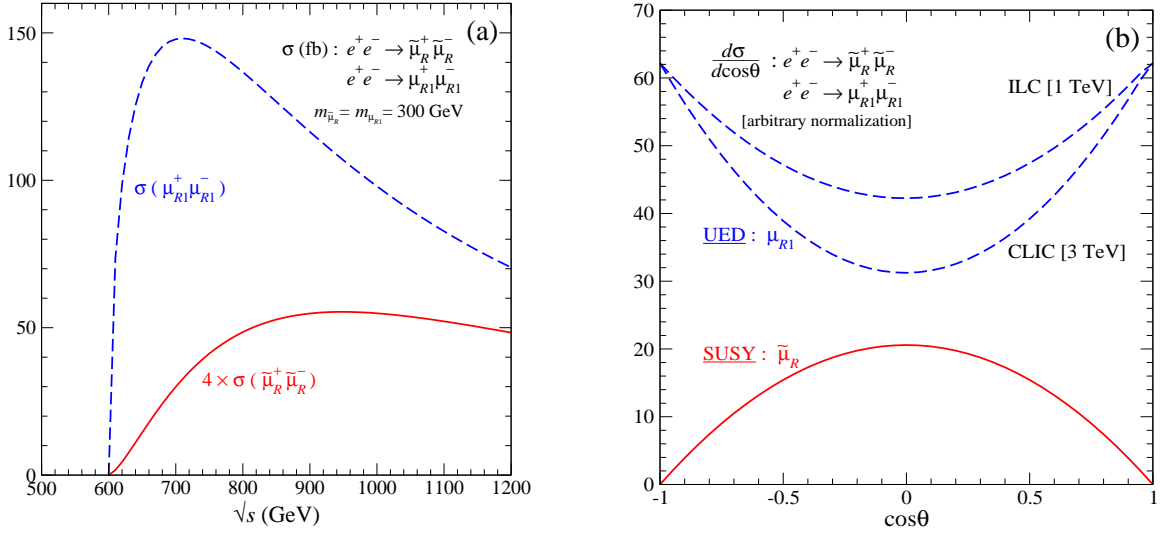

\begin{center}
\includegraphics[height=7.cm,width=7.cm,angle=0]{smuon_threshold.eps}
\hskip 1.2cm
\includegraphics[height=7.cm,width=7.cm,angle=0]{smuon_angle.eps}
\end{center}
\vskip -0.5cm \caption{\it (a) The threshold excitation for smuons; and
             (b) the angular distribution in $e^+e^-\to\tilde{\mu}^+_R
             \tilde{\mu}^-_R$ [arbitrary normalization]. The predictions
             for smuons are compared
             with the corresponding observables for the first KK
             excitation $\mu_{R1}^\pm$ in $e^+e^-\to {\mu}^+_{R1}
             {\mu}^-_{R1}$.}
\label{fig:smuon_muon1}
\end{figure}

The production of spin-0 particles in $e^+e^-$ annihilation is thus
described by two characteristics:
\begin{eqnarray}
&& \#1\quad \mbox{threshold excitation} \quad \sim \ \ \beta^3
  \label{eq:rule_1}\\
&& \#2\quad \mbox{angular distribution} \quad\sim \ \ \sin^2\theta
  \label{eq:rule_2}
\end{eqnarray}
The threshold excitation for smuons and the angular distribution are
illustrated in Fig.$\,$\ref{fig:smuon_muon1}(a) and (b), respectively.
The $\tilde{\mu}_R$ mass is chosen $m_{\tilde{\mu}_R}=300$ GeV.
In the following subsections it will be proven that the
\underline{angular distribution} is characteristic indeed for spinless
particles and that it can be measured with great accuracy
in $e^+e^-$ collisions.\s

%

\subsection{KK Excited States \boldmath{$\mu^\pm_1$} in UED}

The minimal UED version with one universal extra space dimension,
which is compactified on the orbifold $S_1/Z_2$, generates a tower of
spin-1/2 KK states over each L-type and R-type fermion in the Standard
Model (SM) without generating additional zero modes \cite{appelquist}.
Focussing on the R-type states in analogy to the previous subsection, the
process for the production of a pair of the first muonic KK excitation,
\begin{eqnarray}
e^+e^-\ \ \rightarrow\ \ \mu^+_{R1}\, \mu^-_{R1}
\end{eqnarray}
is described by the same diagram Fig.$\,$\ref{fig:diagram1}(a).
$\mu^\pm_{R1}$ is a massive Dirac fermion carrying the electroweak charges of
$\mu^\pm_R$ but coupling only through vector currents to the electroweak gauge
bosons $\gamma$ and $Z$. Thus the generalized charges are identical with
Eqs.$\,$(\ref{eq:smuon_Q_R}) and (\ref{eq:smuon_Q_L}).
It decays into the $\mu^\pm_R \gamma_1$ channel where $\gamma_1$ is the lightest
and stable excited gauge boson generally identified with
the U(1) boson.\footnote{For the sake of simplicity we ignore the electroweak
mixing of the KK excited gauge bosons and identify $W^3_1=Z_1$ and
$B_1=\gamma_1$ in the SU(2) and U(1) gauge boson sectors; the mixing is
suppressed if the KK scale is much larger than the electroweak
scale \cite{Cheng:2002iz}.}\s

The total cross section and the angular distribution are markedly
different from the supersymmetric case:
\begin{eqnarray}
\sigma[e^+e^-\to\mu^+_{R1}\mu^-_{R1}] &=&
   \frac{2\pi\alpha^2}{3s}\,\beta\,\frac{(3-\beta^2)}{2}
   \left[Q^2_L+Q^2_R\right]\\[2mm]
\frac{1}{\sigma}\frac{d\sigma}{d\cos\theta}\,
   [e^+e^-\to\mu^+_{R1}\mu^-_{R1}]
    &=& \ \ \frac{3}{8}\frac{2}{(3 - \beta^2)}
   \left[1+\cos^2\theta+(1-\beta^2)\sin^2\theta \right]
\label{eq:KK_muon_angular_distribution}
\end{eqnarray}
[Because of the vector character of the electroweak $\mu_{R1}$ currents the
distribution is symmetric in forward and backward direction.]
The $\mu_{R1}$ pair is produced in an $S$-wave with non-vanishing amplitude
at the origin. The onset of the excitation curve is therefore suppressed
only by the phase space factor $\sim\beta$. The angular distribution
is familiar from QED, being isotropic at threshold and evolving
to the transverse-polarization term $(1+\cos^2\theta)$ asymptotically.\s

For high energies the total cross section
\begin{equation}
\sigma\ \ \to\ \ \frac{20 \pi \alpha^2}{24 c_W^4}\,\frac{1}{s} \quad
     \mbox{for}\quad  s\ \ \rightarrow \ \ \infty
\end{equation}
scales in the same way as the $\tilde{\mu}^+ \tilde{\mu}^-$ cross section but with
a coefficient 4 times as large, as familiar from QED processes. \s

Choosing for illustration a $\mu^\pm_{R1}$ mass of 300 GeV, a value at the lower
limit of the experimentally allowed range \cite{appelquist}, the onset of the
cross section and the angular distribution are displayed in
Fig.$\,$\ref{fig:smuon_muon1}(a) and (b). The center-of-mass energy is set
to $\sqrt{s}=1$  TeV for ILC \cite{ILC} and 3 TeV for CLIC \cite{CLIC} in the
second figure. Both characteristics are markedly different from
supersymmetric theories; see also Ref.~\cite{battaglia}.
In contrast to scalar smuons the onset of the excitation is vertical, proportional
to the velocity $\beta\sim [s-4 m^2_{\mu_{R1}}]^{1/2}$ as familiar for spin-1/2
particles. Also the angular distributions for scalar smuons and fermionic
KK muon states are distinctly different.  \s

Already from the totally inclusive measurement of the production cross section
no more confusion can arise between supersymmetric theories and theories of
universal extra space dimensions.\s

\subsection{General Analysis}

Though the difference between the characteristics in the production
of supersymmetric scalar particles and KK excited fermions can be
exploited to rule out the false theory experimentally, we should explore
nevertheless whether the conditions (\ref{eq:rule_1}) and/or (\ref{eq:rule_2})
are not only necessary but also sufficient to single out the scalar
solution.\s

The general analysis is most transparent if performed in the helicity
formalism. In the process
\begin{eqnarray}
e^+e^-\ \ \to \ \ F^J_{\lambda_1} \bar{F}^J_{\lambda_2}
\label{eq:general_process}
\end{eqnarray}
for pointlike particles\footnote{Though only string theories are known to
be consistent for interacting particles with $J>2$, weakly interacting field
theories can nevertheless be studied in approaches as formulated in
Ref.~\cite{ferrara}.} $F^J$ and $\bar{F}^J$ with spin $J$ and helicities
$\lambda_1$ and $\lambda_2$ [either half-integer or integer], and mediated
by $s$-channel $\gamma$ and $Z$ exchange, the helicities
of the electron and positron in the initial state are coupled to a vector
with spin $m=\pm 1$ along the beam axis. The right-hand side of the
diagram in Fig.$\,$\ref{fig:diagram1}(a) may then be interpreted as the decay
of a virtual vector boson with polarization $m=\pm 1$ to the $F^J\bar{F}^J$ pair.
If the flight axis of the $F^J$ particles includes the angle $\theta$
with the vector boson polarization axis [identical with the $e^\pm$ beam axis],
the decay amplitude may be expressed in term of the helicity
amplitude ${\cal T}_{\lambda_1\lambda_2}$,
\begin{eqnarray}
\langle F^J_{\lambda_1}\bar{F}^J_{\lambda_2};\theta|V_m\rangle
  = \sqrt{2s}\, g_V\, d^1_{m,\lambda'}(\theta)\,{\cal T}_{\lambda_1\lambda_2}
   \quad \mbox{with} \quad  \lambda'=\lambda_1-\lambda_2
\end{eqnarray}
The angular dependence is in total encoded in the Wigner $d$ function
while the reduced helicity amplitudes ${\cal T}_{\lambda_1\lambda_2}$ are
independent of $m$. The value of $\lambda'$ is
restricted to $\lambda'=0$ and $\pm 1$. The totally inclusive cross section
can be expressed in the form
\begin{eqnarray}
\sigma = \frac{\pi\alpha^2}{3s}\,\beta\,\left(Q^2_L+Q^2_R\right)
          \left[\sum_\lambda
                  \left(|{\cal T}_{\lambda,\lambda-1}|^2
                       +|{\cal T}_{\lambda,\lambda+1}|^2\right)
                       +\sum_\lambda
                        |{\cal T}_{\lambda\lambda}|^2\right]
\end{eqnarray}
and the forward-backward symmetric part of the differential cross section
analogously as
\begin{eqnarray}
\frac{d \sigma'}{d \cos\theta} = \frac{\pi\alpha^2}{4s}
       \beta\left(Q^2_L+Q^2_R\right)
       \left[\frac{1+\cos^2\theta}{2} \sum_\lambda
                  \left(|{\cal T}_{\lambda,\lambda-1}|^2
                       +|{\cal T}_{\lambda,\lambda+1}|^2\right)
                       +\sin^2\theta\sum_\lambda
                        |{\cal T}_{\lambda\lambda}|^2\right]
\end{eqnarray}
summed over the helicities of the outgoing $F^J$ and $\bar{F}^J$ particles.
To evaluate these expressions, the two cases in which $F^J$ is either
fermionic or bosonic must be distinguished.\\[1.5mm]

\noindent
{\bf (a)} \underline{Fermionic spectrum} $F[J=1/2, 3/2, \ldots]$:\\
For pointlike theories in which the fermions carry electric and weak monopole
charges the associated vector current includes the basic
component \cite{ferrara}
\begin{eqnarray}
j^0_\mu = \bar{\psi}_{\alpha_1\cdots\alpha_n}\gamma_\mu
             \psi^{\alpha_1\cdots\alpha_n}
\label{eq:no_derivative_current}
\end{eqnarray}
built by the $J=n+1/2$ spinor-tensor wave-function $\psi^{\alpha_1,\ldots,
\alpha_n}$ \cite{wave_functions}. In analogy to the Dirac spin 1/2 case, the current
(\ref{eq:no_derivative_current}) can be decomposed, {\it cf.}
Refs.~\cite{ferrara,paschos}, into an electric current $j^e_\mu$ and a magnetic
current $j^m_\mu$ as $j_\mu=j^e_\mu+j^m_\mu$ with\footnote{Demanding
asymptotic unitarity at high energies, additional terms must be included in the
basic Lagrangian \cite{Singh:1974qz} which alter the gyromagnetic ratio
from $g=J^{-1}$ universally to $g=2$ \cite{ferrara}, {\it i.e.} the coefficient
in front of Eq.$\,$(\ref{eq:magnetic_term}) may be adjusted accordingly.
Universality of the value $g=2$ beyond the Dirac theory is a well-known prediction
of non-abelian gauge theories.}
\begin{eqnarray}
&& j^e_\mu = \frac{1}{2m} \bar{\psi}_{\alpha_1,\ldots,\alpha_n}\,\,
   i\, \overleftrightarrow{\partial_\mu}\,\, \psi^{\alpha_1,\ldots,\alpha_n}
   \label{eq:electric_term}\\[2mm]
&& j^m_\mu =\frac{1}{2m} \partial^\nu
  \left( \bar{\psi}_{\alpha_1,\ldots,\alpha_n}\,\,
  \sigma_{\mu\nu}\,\, \psi^{\alpha_1,\ldots,\alpha_n} \right)
  \label{eq:magnetic_term}
\end{eqnarray}
reducing to electric monopole and magnetic dipole currents in the
non-relativistic limit.\s

Both the electric current $j^e_\mu$ and the magnetic current $j^m_\mu$
give rise to diagonal reduced helicity amplitudes through
spin-0/$D$-wave
and spin-1/$S$-wave interactions, respectively; apart from overall
coefficients,
\begin{eqnarray}
&& {\cal T}^e_{\lambda\lambda}\, =\, \phantom{-}\frac{\gamma\beta^2}{\sqrt{2}}
   \left[ \frac{(J+\lambda)}{2J}\, Q^{J-1/2}_{\lambda-1/2}(\gamma)
         -\frac{(J-\lambda)}{2J}\, Q^{J-1/2}_{\lambda+1/2}(\gamma)\right] \\
&& {\cal T}^m_{\lambda\lambda}\, =\,\,
   -\frac{\gamma}{\sqrt{2}} \left[ \frac{(J+\lambda)}{2J}\,
          Q^{J-1/2}_{\lambda-1/2}(\gamma)
         -\frac{(J-\lambda)}{2J}\, Q^{J-1/2}_{\lambda+1/2}(\gamma)\right]
\end{eqnarray}
Only the magnetic dipole current $j^m_\mu$ generates non-diagonal reduced
helicity amplitudes,
\begin{eqnarray}
{\cal T}^m_{\lambda,\lambda\pm 1}\, =\,\mp
  \frac{\sqrt{(J\mp \lambda)(J\pm\lambda +1)}}{2J}\,\,
  Q^{J-1/2}_{\lambda\pm 1/2} (\gamma)
\label{eq:fermion_off_diagonal}
\end{eqnarray}
Here, $\gamma= \sqrt{s}/2m$ is the Lorentz boost factor of the final-state
particle and the energy-dependent functions $Q^N_n(\gamma)$ ($|n|\leq N$ for
integral $N$ and $n$) are defined as
\begin{eqnarray}
Q^N_n(\gamma) &=& \frac{2^N (N+n)!\, (N-n)!}{(2N)!}\,
    \sum_{\lambda_1=\pm 1, 0}\cdots\sum_{\lambda_N=\pm 1, 0}\,
    \delta_{n,\lambda_1+\cdots+\lambda_N}\,
    \prod_{i=1}^N\frac{(2\gamma^2\delta_{\lambda_i 0}-1)}{
                       (1+\lambda_i)!\, (1-\lambda_i)!}\\
Q^{\,0}_0(\gamma) &=& 1\quad \mbox{and}\quad
Q^N_n(\gamma)\, =\, 0\quad \mbox{for}\quad |n|> N\geq 0
\end{eqnarray}
We note that the non-diagonal reduced helicity amplitudes
${\cal T}_{\lambda,\lambda\pm 1} = {\cal T}^m_{\lambda,\lambda\pm 1}$ are
non-vanishing for any energy. As a result, the term
$\sum_\lambda\left(|{\cal T}_{\lambda,\lambda-1}|^2
+|{\cal T}_{\lambda,\lambda+1}|^2\right)$
never vanishes, leading to a cross section that rises $\sim\beta$ at
the threshold and contributing with a term $\sim (1+\cos^2\theta)$
to the angular distribution. Both elements differ clearly from the
production of scalar particles. We therefore conclude that scalar
spin-0 particles in supersymmetric theories carrying muon-type
charges can never be confused by fermionic charged particles.\\[1.5mm]

\noindent
{\bf (b)} \underline{Bosonic spectrum} $F[J=1,2,\ldots]$:\\
Restricting ourselves to CP-invariant theories, the electric and weak
monopole charge term of any integer spin $J$ tensor-field
$\varphi^{\alpha_1..\alpha_J}$ is accounted for by the current element
\begin{eqnarray}
j^e_\mu \,=\, i\,\, \varphi^*_{\alpha_1\cdots\alpha_J}\,\,
          \overleftrightarrow{\partial_\mu}\,\,
          \varphi^{\alpha_1\cdots\alpha_J}
\label{eq:integer_spin_monopole}
\end{eqnarray}
It leads to $P$-wave production of the boson pair with the
reduced helicity amplitude
\begin{eqnarray}
{\cal T}^e_{\lambda\lambda}\, =\, -\frac{\beta}{\sqrt{2}}\,  Q^J_{\lambda} (\gamma)
\end{eqnarray}
Since the wave-function vanishes at the origin, the total production cross
section rises $\sim \beta^3$ at the threshold. Thus, opposite to wide-spread
belief, the onset of the excitation
curve near threshold does not discriminate the spin 0 particle from higher
integer spin $J=1,2,\ldots$ particles.\s

However, coupling the electroweak vector fields to the spin $J$ fields in a
consistent way \cite{ferrara}, a non--vanishing magnetic dipole moment is
generated for all particles with spin $>0$. The non--zero magnetic current, which
is proportional to\footnote{As before, asymptotic unitarity \cite{ferrara}
modifies the coefficient of this current such that the gyromagnetic ratio
is shifted again from $g=J^{-1}$ to $g=2$.}
\begin{eqnarray}
j^m_{\mu} = -i\,\partial^\nu(\varphi_{[\mu}^{*\alpha_2\ldots\alpha_J}
                \,\varphi_{\nu]\alpha_2\ldots\alpha_J})
\end{eqnarray}
gives rise to non-vanishing off-diagonal helicity amplitudes which can be written,
apart from an overall coefficient, as
\begin{eqnarray}
{\cal T}^m_{\lambda,\lambda\pm 1}
  \,=\, -
\gamma\beta \, \frac{\sqrt{(J\mp\lambda)(J\pm\lambda +1)}}{2J (2J-1)}\,
     \left[(J\pm\lambda)\, Q^{J-1}_{\lambda} (\gamma)
          +(J\mp\lambda-1)\, Q^{J-1}_{\lambda\pm 1}(\gamma)\right]
\label{eq:boson_off_diagonal}
\end{eqnarray}
The $P$-wave behavior near the threshold is reflected in the coefficient
$\beta$. The non-vanishing helicity amplitude ${\cal T}_{\lambda,\lambda\pm 1}
={\cal T}^m_{\lambda,\lambda\pm 1}$
for $J>0$ is in apparent contrast to  spin-0 scalars for which these
amplitudes must vanish. Thus opposite to scalar production, higher spin
$J=1,2,..$ production will generate an additional term
$\sim (1+\cos^2\theta)$ in the angular distribution, non-vanishing
in the forward and backward direction. Thus the analysis of the
angular distributions signals the zero-spin of the smuons
unambiguously.\s

Complementary to this theory-based argument on production properties,
{\it i.e.} the onset of the excitation curve and the angular distribution,
decay characteristics can also be exploited to supplement the analysis.
The presence of the off-diagonal helicity amplitudes,
Eq.$\,$(\ref{eq:fermion_off_diagonal}) for fermions and
Eq.$\,$(\ref{eq:boson_off_diagonal}) for bosons, implies
that the final state particles $F^J$ and $\bar{F}^J$ should be polarized
when electron and/or positron beams are polarized.
In addition, the relation $\lambda_1-\lambda_2=\pm 1$ for the non-vanishing
off-diagonal helicities of the $F^J_{\lambda_1}\bar{F}^J_{\lambda_2}$ pair
produced via $\gamma, Z$ exchange should lead to interesting polarization
correlations which can be observed through the correlated decays of
$F^J_{\lambda_1}$ and $\bar{F}^J_{\lambda_2}$.
In this case, the polar-angle distribution of the decay particles in
the $F^J$ rest frame is described by the Wigner $d$ function,
\begin{eqnarray}
\mathcal{D}\left[F^J_\lambda \rightarrow a_{\sigma_1} b_{\sigma_2}\right]
\,\,\sim\,\,
d^J_{\lambda \sigma}(\theta^*) \quad\mbox{with}\quad
\sigma=\sigma_1-\sigma_2
\end{eqnarray}
where $\theta^*$ denotes the polar angle between the $F^J$ flight direction and
the $ab$ axis in the $F^J$ rest frame. This configuration is realized
in the dominant $\mu\tilde{\chi}^0_1$ decay channel of $\tilde{\mu}_R$.
For scalar smuons $d^0_{00}(\theta^*)$ does not depend on $\theta^*$. For
$J>0$, however, even if the final-state polarizations $\sigma_1,\sigma_2$ are
summed over, the angular distribution is always non-trivial when the sum of
two final-state daughter spins, $j_a$ and $j_b$, is less than the spin $J$ of
the parent. In the opposite case, $j_a+j_b\geq J$, parity-violating
decays in general guarantee non-trivial angular dependence\footnote{Only in
exceptional cases, like $\tau\to \nu_\tau a_1$ with
$m_\tau\simeq \sqrt{2}m_{a_1}$, P-violating decays cannot be used as
polarization analyzer.}; only in
parity-preserving cases the decay distribution might be independent of the
$F^J$ polarization. However, if the final-state polarizations $\sigma_1,\sigma_2$
are measured, the $\theta^*$ dependence of $d^J_{\lambda\sigma}(\theta^*)$
for $J > 0$ is always non-trivial, whatever values are taken for $|\lambda|$
and $|\sigma|\leq J$.
After squaring the decay amplitude $\mathcal{D}$, the spin $J$ can be determined
by projecting out the maximum spin index $2J$ from the decay angular
distributions.\s

Therefore the analysis of the smuon decay
distributions provides us with an alternative model-independent
method for the determination of the zero smuon spin.\s

\subsection{Reconstructing the Event Axis}

{\bf (a)} \underline{$\smur^+\smur^-$ in supersymmetry}:\\
The measurement of the cross section for smuon
pair production $\tilde{\mu}^+_R\tilde{\mu}^-_R$
can be carried out by identifying acoplanar $\mu^+\mu^-$ pairs
[with respect to the $e^\pm$ beam axis] accompanied by large missing energy:
\begin{eqnarray}
  e^+e^-\ \ \rightarrow\ \ \tilde{\mu}^+_R\,\tilde{\mu}^-_R\ \
  \rightarrow\ \ (\mu^+\tilde{\chi}^0_1) \;  (\mu^-
  \tilde{\chi}^0_1)\ \ \to \ \ \mu^+\mu^-\not\!\!{E}  \
\end{eqnarray}
The analysis is model-independent and it provides unambiguously the
onset of the excitation curve $\sim\beta^3$ near threshold.\s

The construction of the production angle $\theta$
is illustrated for the event topology in Fig.$\,$\ref{fig:sketch}.
For very high energy $\sqrt{s}\gg m_{\tilde{\mu}_R}$ the
flight direction of the daughter particles $\mu^\pm$'s can be approximated by
the flight direction of the parent particle \cite{battaglia} and the
dilution due to the decay kinematics is small.
However, at medium ILC energies the dilution increases, and the reconstruction
of the $\tilde{\mu}^\pm_R$ flight direction provides more accurate
results on the angular distribution of the smuon pairs \cite{martyn}.
If all particle masses are known,
the magnitude of the particle momenta is calculable and
the relative orientation of
the momentum vectors of $\mu^\pm$ and $\smur^\pm$ is fixed
by the two-body decay kinematics.
The opening angles $\alpha_\pm$ between the visible
$\mu^\pm$ tracks and the parent $\tilde{\mu}^\pm_R$ particles can
be determined from the relation
\begin{eqnarray}
  m^2_{\mu^\pm_R} - m^2_{\tilde{\chi}^0_1} = \sqrt{s}E_{\mu^\pm}
  (1-\beta_{\tilde{\mu}^\pm_R} \cos\alpha_\pm)
  \label{eq:cos_alpha}
\end{eqnarray}
The angles $\alpha_\pm$ define two cones about the $\mu^+$ and $\mu^-$
axes which intersect in two lines
-- the true $\tilde{\mu}^\pm_R$ flight direction and a false direction.
True and false solutions are mirrored on the plane spanned by the $\mu^+$
and $\mu^-$ flight directions. Thus
the flight direction can be reconstructed up to a 2-fold ambiguity.\s

\begin{figure}[htb]
\begin{center}
{\color{black}
\begin{picture}(200,170)(0,0)
\Text(5,78)[r]{\boldmath{$e^-$}}
\Text(195,78)[l]{\boldmath{$e^+$}}
\ArrowLine(10,75)(95,75)
\ArrowLine(190,75)(105,75)
\Vertex(100,75){2.5}
\SetColor{Red}
\LongArrow(100,75)(170,145)
\LongArrow(100,75)(30,5)
\SetColor{Blue}
\LongArrow(100,75)(110,135)
\LongArrow(100,75)(20,40)
\SetColor{Black}
\Text(118,105)[c]{\color{black} $\alpha_-$}
\Text(63,48)[c]{\color{black} $\alpha_+$}
\CArc(100,75)(25,45,82)
\CArc(100,75)(25,205,225)
\SetColor{Blue}
\DashLine(110,135)(170,145){2}
\DashLine(30,5)(20,40){3}
\SetColor{Black}
\Text(55,10)[]{\color{red} $\tilde{\mu}^+_R$}
\Text(175,130)[]{\color{red} $\tilde{\mu}^-_R$}
\Text(105,120)[r]{\color{blue} $\mu^-$}
\Text(40,55)[r]{\color{blue} $\mu^+$}
\Text(135,152)[c]{\color{blue} $\tilde{\chi}^0_1$}
\Text(10,20)[c]{\color{blue} $\tilde{\chi}^0_1$}
\Text(137,90)[l]{\large\color{red} $\theta$}
\SetColor{Red}
\LongArrowArc(100,75)(30,0,43)
\end{picture}
}
\end{center}
\caption{\it Event topology of the reaction $\ee\to\smur^+\smur^-\to
             \mu^+\nt_1\,\mu^-\nt_1$}
\label{fig:sketch}
\end{figure}

\begin{figure}[ht!]
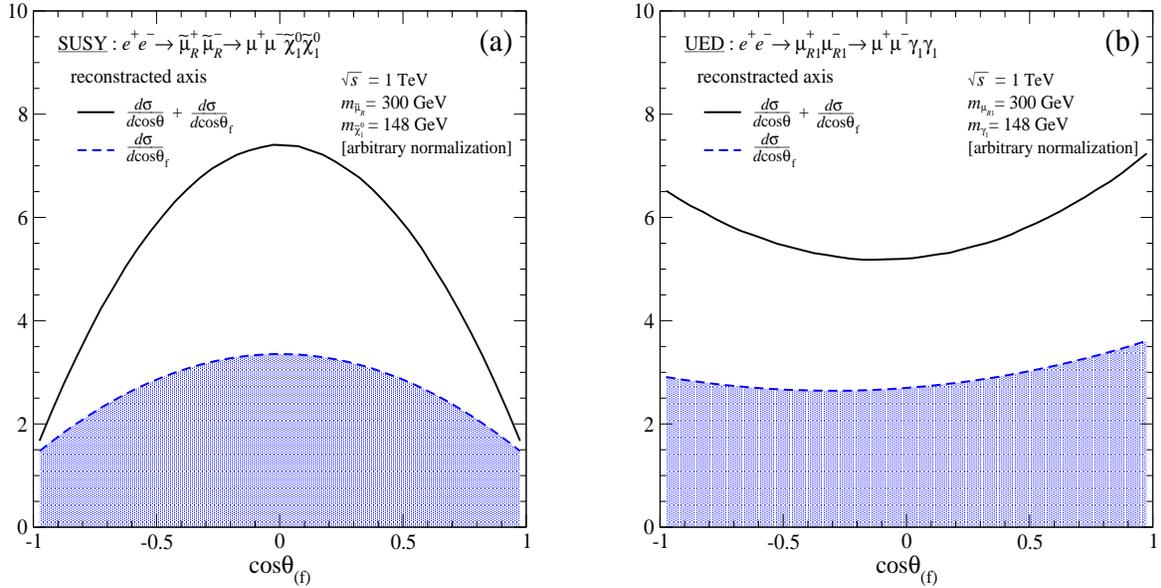

\begin{center}
\includegraphics[height=7.8cm,width=7.cm,angle=0]{wrngSUSYaxis.eps}
\hskip 1.2cm
\includegraphics[height=7.8cm,width=7.cm,angle=0]{wrngUEDaxis.eps}
\end{center}
\caption{\it (a) The angular distribution of the false
       reconstructed axis (blue shaded area) and of the observable sum
       of true and false axes for smuon pair production in SUSY (black);
       (b) The same for $\mu^+_{R1}\mu^-_{R1}$ pair production in UED.
       [The distribution of the false axis is slightly forward-backward
       asymmetric due to the
       parity violating $\mu^\pm_{R1} \to \mu^\pm \gamma_1$ decay.]}
\label{fig:figwrng}
\end{figure}

The characteristics of the angle $\theta_{ft}$ between the false and the true
axis can easily be illustrated. If the decay planes of $\tilde{\mu}^+_R$ and
$\tilde{\mu}^-_R$ coincide, the production axis is located in the common plane
and the false axis coincides with the true axis. Rotating one of the two planes
away from the other by an azimuthal angle $\phi$, the angle $\theta_{ft}$ between
the false and the true axis is related to $\phi$ and to the boosts
$\gamma_\pm =\gamma (\cos\alpha^*_\pm+\beta)/\sin\alpha^*_\pm$ with $\alpha^*_\pm$
being the $\tilde{\mu}^\pm_R\to \mu^\pm$ decay angle in the $\tilde{\mu}^\pm_R$
rest frame with respect to the flight direction in the laboratory frame:
\begin{eqnarray}
\cos\theta_{ft} = 1 -\frac{2 \sin^2\phi}{
           \gamma^2_++\gamma^2_-+2\gamma_+\gamma_-\cos\phi+\sin^2\phi}
\end{eqnarray}
For high energies the maximum opening angle reduces effectively
to $\theta_{ft} \lesssim {\rm min}(1/|\gamma_+|,1/|\gamma_-|) = O(1/\gamma)$ and
approaches zero asymptotically when the two axes coincide. Quite generally,
as a result of the Jacobian root singularity in the relation between
$\cos\theta_{ft}$ and $\phi$, the false solutions tend to accumulate slightly
near the true axis for all energies. In total, the angular distribution of the
false axis with respect to the true axis is given by
\begin{eqnarray}
\frac{dN}{d\cos\theta_{ft}}\, = \, \frac{\sqrt{2}}{\sqrt{1-\cos\theta_{ft}}}\cdot
  \frac{F\left[\beta\cos(\theta_{ft}/2)\right]}{\gamma^4\beta^4
  \left(1-\cos\theta_{ft} +2/\gamma^2\beta^2\right)^2\,}
\end{eqnarray}
with $F[0]=1$ at threshold and $F[\beta]\sim \gamma$ for $\beta\to 1$.
The decrease of the coefficient $\sim 1/\gamma^4$ is compensated by the effective
narrowing of $\theta_{ft}$ in the denominator and by the increase of the
function $F$ for rising energy. Thus, the false axis is
trailed by the true axis, mildly at low energies and tightly at high energies.
Though the distribution of the false axis is flattened at low to medium energies
compared with the original distribution of the true axis, the characteristic
features are reflected qualitatively, nevertheless,
{\it cf.} Fig.$\,$\ref{fig:figwrng}. For our theoretical investigation throughout
the paper, the R-type slepton mass $m_{\ell_R}=300$ GeV and sneutrino mass
$m_{\tilde{\nu}}=365$ GeV are used and the chargino/neutralino mass spectra and the
mixing elements are derived from the SUSY Lagrangian parameters $M_2 =$ 300 GeV,
$M_1 =$ 150 GeV, $\mu =$ 500 GeV and $\tan\beta = 10$. This parameter set includes
the lighter chargino mass $m_{\tilde{\chi}^\pm_1}=286$ GeV  and to
the two lowest neutralino masses $m_{\tilde{\chi}^0_{1/2}}=148/286$ GeV.\s\s

Experimentally, the absolute orientation in space is operationally obtained by
rotating the two $\smur^\pm$ vectors around the $\mu^\pm$ axes against each other
until they are aligned back-to-back in opposite directions.
The flattened false-axis distribution can be subtracted on the basis
of Monte Carlo simulations.
For a fraction of events the production angle cannot be reconstructed,
which in most cases is due to large initial state radiation and/or
beamstrahlung reducing the nominal center of mass energy considerably.
In addition losses occur due to measurement errors of the final particles.
It is also important to note that background events, if they can be
reconstructed at all under the wrong (mass) hypothesis, usually produce flat
angular distributions and can thus be easily subtracted.\s\s

\noindent
{\bf (b)} \underline{$\mu^+_{R1}\mu^-_{R1}$ in UED}:\\
As proven in the previous subsection, the experimental observation of the
$\sin^2\theta$ law determines the spin-0 character of new particles carrying
non-electron fermion numbers, {\it i.e.} smuons, squarks, {\it etc},
unambiguously. This general conclusion can be illuminated by analyzing
the spin-1/2 angular distribution of the UED KK excitation $\mu^\pm_{R1}$. The
[non-normalized] angular distribution of $\mu^+_{R1}\mu^-_{R1}$ pairs in
$e^+e^-$ collisions is described in Eq.$\,$(\ref{eq:KK_muon_angular_distribution}):
$N(\theta) = 1 + \cos^2\theta \, + \, (1 - \beta^2)\,\sin^2\theta$.
The decay $\mu^\pm_{R1} \to \mu^\pm \gamma_1$ in the $\mu^\pm_1$ rest frame is
governed by the right-handed coupling between the two leptons so that $\mu^-$,
including the angle $\theta_1$ with the $\mu^-_{R1}$ polarization vector,
is preferentially emitted in the direction parallel to the $\mu_{R1}^-$
polarization vector, $D(\theta_1) = (1+\kappa_{R1} \cos\theta_1)^2$
with $\kappa_{R1}=(m^2_{\mu_{R1}}-2 m^2_{\gamma 1})/
(m^2_{\mu_{R1}}+2 m^2_{\gamma 1})\sim 1/3$.
The opposite rule applies to $\mu_{R1}^+$ decays. \s

Properly including the correlations among the two decay pairs, the predictions
for the distributions of the true production axis, $N(\theta)$, and the false
axis, $N_f(\theta)$, are displayed in Fig.$\,$\ref{fig:figwrng}(b). After
subtracting the distribution of the false axis from the sum, the distribution of
the true axis is markedly different from the distribution of the smuon polar
production angle in Fig.$\,$\ref{fig:figwrng}(a). In particular, the production
of spin-1/2 KK muons populates the forward and backward directions in contrast
to spin-0 smuons. \s

This method can be applied quite generally in $e^+e^-$ annihilation through
$\gamma$ and $Z$ exchange for any given theory. For the set $\{c_1, c_2\}$ of
coefficients in the true angular distribution
\begin{eqnarray}
N(\theta)\ \ \sim\ \ c_1\,[1+\cos^2\theta] + c_2\, \sin^2\theta
\end{eqnarray}
the false distribution $N_f(\theta)$ can be generated unambiguously.
Comparing the sum of the distributions of the experimentally reconstructed
true and false events with $N+N_f$, the ratio of the two coefficients
$c_2/c_1$ can be fitted by using template methods. The fit will only be
acceptable if at the same time the helicity structure of the decay vertex
is chosen correctly. \s

\subsection{Experimental Analysis}
\label{sec:experimental_analysis}

\noindent {\bf (a)} \underline{Sparticle spectrum}:
In order to perform a realistic simulation of signal and background
processes a sparticle spectrum is calculated using the program
{\sc Isajet~7.74}~\cite{isajet}.
The SUSY point described above can well be embedded in a mSUGRA
scenario\footnote{The particle masses corresponding to this reference
point differ slightly at a level of a few GeV from the previously
adopted masses in the theoretical illustrations.}
with the parameters
$m_0=265\;\GeV, M_{1/2}=375\;\GeV , \tan\beta=10, A_0=0,$ and
${\rm sign}\,\mu = +1$, corresponding to the Lagrangian parameters
$M_1= 156\;\GeV, M_2=291\;\GeV,  \mu=488\;\GeV$.
The masses of the R/L-type smuons/selectrons, electron sneutrino,
lighter chargino and two lightest neutralinos accessible at a 1~TeV ILC
and relevant for the present experimental study are listed in
Table~\ref{tab:susyspectrum}.\s\s

\renewcommand{\arraystretch}{1.1}
\begin{table}[htb] \centering
  \caption{\it
    Spectrum of sleptons, charginos and neutralinos in the SUSY scenario
    ($M_1= 156\;\GeV, M_2=291\;\GeV,  \mu=488\;\GeV$).}
  \label{tab:susyspectrum}
  \vspace{3mm}
  \begin{tabular}{ l c c l c}
    \hdick
    \phantom{xxxx} & \phantom{xxxxx} &
    \phantom{xxxx} & \phantom{xxxxx} \\ [-3.ex]
    $\sell$     & m [GeV] && $\ch$       & m [GeV]
    \\[.5ex] \hdick
    $\ser/\smur  $     & 302   && $\ch_1^\pm$ & 285\\
    $\sel/\smul  $     & 369   && $\ch_1^\pm$ & 510\\
    $\snu_e/\snu_\mu$  & 359   && $\nt_1 $    & 152\\
    $\tilde{\tau}_1$   & 297   && $\nt_2 $    & 284\\
    $\tilde{\tau}_2$   & 369   && $\nt_3 $    & 493\\
    $\tilde{\nu}_\tau$ & 357   && $\nt_4 $    & 511\\
    \hline
  \end{tabular}
\end{table}

\noindent {\bf (b)} \underline{Event generation}:
Events are generated with the program {\sc Pythia~6.3}~\cite{pythia}
which includes initial and final state QED radiation as well as
beamstrahlung~\cite{circe}.
The experimental simulation is based on the detector proposed
in the {\sc Tesla tdr}~\cite{tdr}
and implemented in the Monte Carlo program {\sc Simdet}~4.02~\cite{simdet}.
The detector requirements are excellent momentum and energy resolution,
good particle identification and full hermetic coverage.
The detector response, resolution and particle reconstruction are treated in a
parametric form. It is further assumed that the ILC can be operated at a
flexible energy up to $\sqrt{s}=1\;\TeV$ and that both lepton beams can be
polarized at a degree of $|\cP_{e^-}|= 0.8$ for electrons and $|\cP_{e^+}|= 0.6$
for positrons. Beam polarization helps to enhance the production rates and to
select the signal but it has no essential influence on the spin analyses of the
distributions under investigation.\s\s

\noindent {\bf (c)} \underline{Event reconstruction}:
The reconstruction of the polar angle of pair production relies on
the knowledge of the masses of the primary and secondary particles.
Based on pure kinematics of two-body decays, like
$\sell^\pm\to\ell^\pm\nt_1$, $\cx^\pm_1\to W^\pm\nt_1$ and $\nt_2\to Z\nt_1$,
both masses can be determined from the energy spectra of
the observable decay particle, see {\it e.g.} Ref.~\cite{Martyn:1999tc}.
Alternatively, the excitation curve can be used to determine the mass
of the primary SUSY particle pair. However, the observable cross section
close to threshold is in general distorted considerably and the theoretical
expectation has to be convoluted with initial state radiation (ISR),
beamstrahlung (BS) and finite width effects. ISR can be rigorously treated
in QED. The BS energy profile depends on the collider operation conditions,
and it can be measured via Bhabha scattering \cite{Bha}; it can also
be calculated for given machine parameters~\cite{circe}. The width of SUSY
particles is calculable within a specific model, but can also be determined in
a simultaneous fit of the excitation curve~\cite{Martyn:1999xc,RC1,RC2}.
It can be safely assumed that the sparticle masses can be measured with
a precision of one permille or better, see Ref.~\cite{Freitas:2004re}.
Such an accuracy is sufficient for the present study to reconstruct the event
kinematics reliably.\s

\subsection{Simulation of \boldmath $\ee\to\smur^+\smur^-$}

The detection of scalar smuons in the reaction $\ee\to\smur^+\smur^-$
with subsequent decays $\smur^\pm\to\mu^\pm\nt_1$ is relatively simple
and clean. The energy spectrum of the decay muon is flat with minimal and
maximal values given by $E_\pm = \sqrt{s}/4\,(1-m_\nt^2/m_\smu^2)\,(1\pm\beta)$
with $\beta=(1-4\,m_\smu^2/s)^{1/2}$.
The event selection criteria are:
(i) two oppositely charged $\mu^\pm$ and nothing else in the detector;
(ii) signed polar angle acceptance
     $-0.90 <  Q_\mu\cos\theta_\mu < 0.75$;
     where $Q_\mu$ is the muon charge and the asymmetric cut
     rejects muons from decays of $W^+W^-$ production;
(iii) acoplanarity angle between the two muons
      $\Delta\Phi_{\mu\mu} < 160^\circ$;
(iv)  the missing momentum vector should point inside the sensitive detector
      $ |\cos\theta_{\vec p_{\rm miss}}| < 0.9$; and
(v)  lepton energy
     within the kinematically allowed boundaries
     $E_- \leq E_\mu \leq E_+$
     (modulo resolutions).
The resulting detection efficiency is typically around $\epsilon\simeq 0.60$.\s

\begin{figure}[tbh!]
\begin{center}
\includegraphics[width=17.cm,height=14.cm,angle=0]{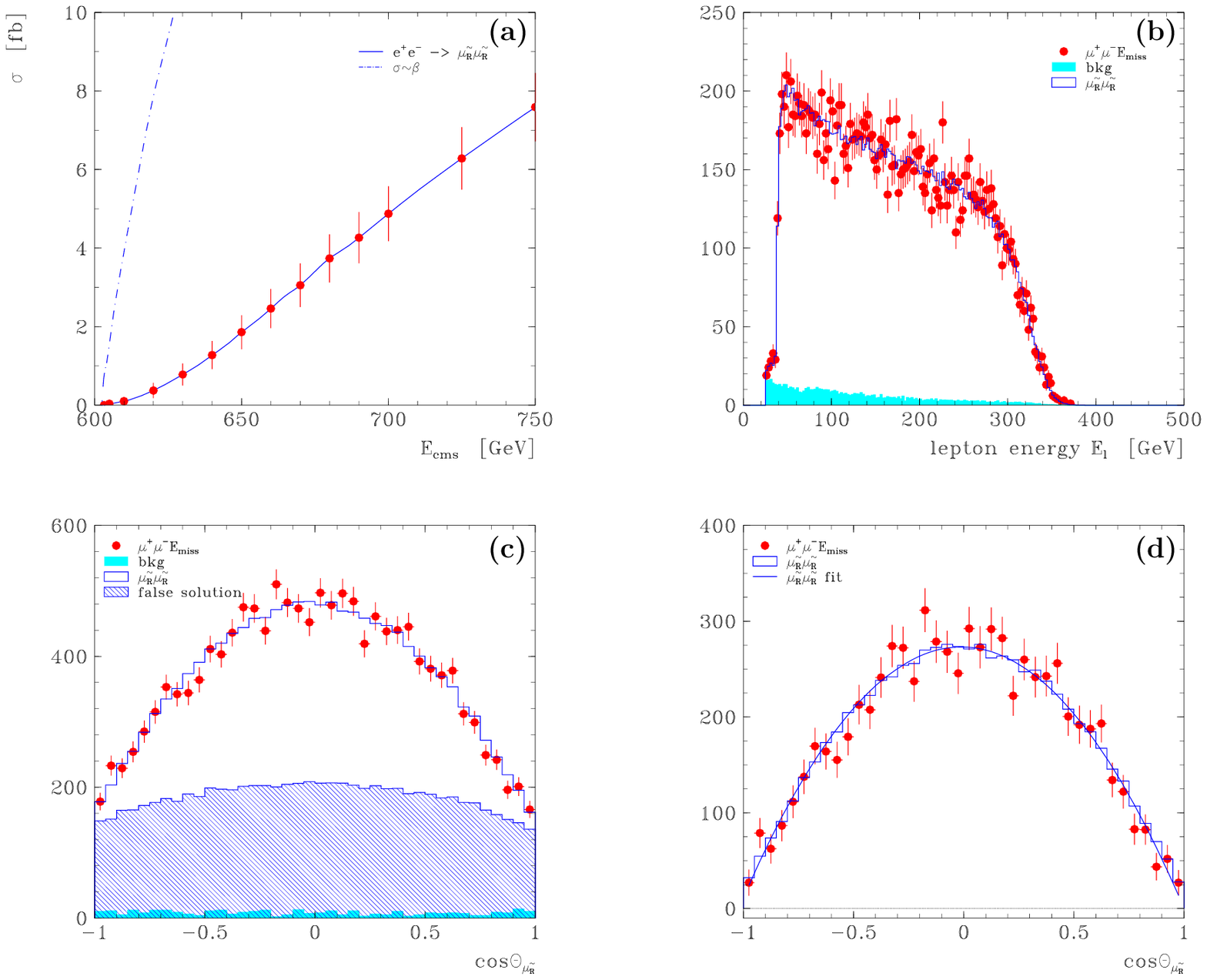}
\end{center}
\vskip -0.5cm
\caption{\it
  (a) The unpolarized cross section of $e^+e^-\to\smur^+\smur^-$
  production close to threshold,
  including QED radiation, beamstrahlung and width effects;
  the statistical errors correspond to $\cL=10\,\fbi$ per point,
  the dash-dotted curve indicates, with the same coefficient,
  the hypothetical dependence $\sigma\sim\beta$
  instead of $\sigma\sim{\beta^3}$;
  (b) energy spectrum $E_\mu$ from $\smur^-\to \mu^-\nt_1$ decays;
  polar angle distribution $\cos\theta_\smur$
  (c) with and (d) without contribution of false solution.
  The simulation for the energy and polar angle distribution
  is based on polarized beams with $(\cP_{e^-},\cP_{e^+}) = (+0.8, -0.6)$
  at $\sqrt{s}=1\,\TeV$ and $\cL=500\,\fbi$.
  The smooth histograms represent high statistics expectations,
  the curve indicates a fit to the cross section
  (\ref{eq:smur_differential_cross_section}).}
\label{fig:figa1}
\end{figure}

The unpolarized cross section as a function of energy close to the
production threshold, including all instrumental effects, is shown in
Fig.$\,$\ref{fig:figa1}(a). The remaining flat background from $W^+W^-$
production (not shown) amounts to about $0.5\;\fb$,
and it can be subtracted by extrapolation from the sideband below.
The excitation curve exhibits a slow rise as expected from the
characteristic dependence $\sigma_{\smur\smur}\sim\beta^3$
explained in Eq.$\,$(\ref{eq:rule_1}).
Such a behavior can be clearly distinguished from a much steeper
hypothetical $S$-wave
dependence $\sigma\sim\beta$, shown as well for comparison.\s

The $\smur^\pm$ angular distribution is investigated in the continuum
choosing an energy of $\sqrt{s} = 1\;\TeV$ and an integrated luminosity
of $\cL=500\,\fbi$. In order to enhance the signal and suppress SUSY
and SM background processes, the beams are assumed to be polarized with
values of $(\cP_{e^-},\cP_{e^+}) = (+0.8, -0.6)$,
resulting in a cross section of $30\;\fb$.\s

The spectrum of the muon energy $E_\mu$ is shown in Fig.$\,$\ref{fig:figa1}(b).
The signal is very clean above the very low background from SUSY and $W^+W^-$
production, which gets reduced further to $\sim 1.3\;\%$ after reconstruction
of the kinematics.
The primary flat energy spectrum, characteristic for a spin 0 particle decay,
is distorted due to acceptance cuts, event selection criteria and
photon radiation.
However, the minimal and maximal endpoint energies $E_\pm$ are clearly
pronounced.
The polar angle distribution of the smuons $\smur$,
including both the correct and false solutions,
is displayed in Fig.$\,$\ref{fig:figa1}(c).
The tiny background gives a flat contribution.
The false solution
exhibits a large pedestal with some enhancement in the central region
reflecting mildly the primary distribution ({\it cf}. $\beta=0.8$).
The ambiguous solution
can be calculated using Monte Carlo simulation and it is
subtracted in Fig.$\,$\ref{fig:figa1}(d).
The expected $\sin^2\theta$ distribution of Eq.$\,$(\ref{eq:rule_2}) is clearly
visible. A fit of the shape to the experimental angular distribution
yields
\begin{eqnarray}
&& \mbox{ }\hskip 1.cm \frac{d \sigma^{\rm exp}}{d \cos\theta}\, \sim\,
  1 + a\cos\theta + b\cos^2\theta \\[2mm]
&& a = -0.020\pm 0.016 \quad  \mbox{and} \quad
   b = -0.979\pm 0.022 \nonumber
\end{eqnarray}
confirming the conjectured forward-backward symmetric $\sin^2\theta$ behavior
of spin-0 smuon production with high precision.\s

\section{SELECTRONS}
\label{sec:selectrons}
\setcounter{equation}{0}

\subsection{Production Channels in \boldmath{$e^+e^-$} Collisions}

In the selectron production process the lepton number can flow
from the initial to the final state. Therefore, besides the $e^+e^-$
annihilation channel mediated by $\gamma, Z$ exchange, {\it cf.}
Fig.$\,$\ref{fig:diagram1}(a),
also $t$-channel exchange of neutralinos, {\it cf.} Fig.$\,$\ref{fig:diagram1}(b),
contributes to some channels. [In $e^-e^-$ collisions selectrons are produced
solely by $t$-channel and $u$-channel exchanges.] Among all these
channels, which are summarized comprehensively in Table~1 of
Ref.~\cite{RC2}, pair production of
$\tilde{e}^+_R\tilde{e}^-_R$ is easiest to control experimentally if,
as realized in many models, the R-type selectron is significantly
lighter than the companion L-type selectron.
Equal-particle channels are also preferred theoretically; their
analysis is most transparent by including the standard annihilation
which is well controlled.\s

Two electron/positron polarization states can generate the
$\tilde{e}^+_R\tilde{e}^-_R$ pair:
\begin{eqnarray}
&& e^+_R\, e^-_L\ \ \to\ \ \tilde{e}^+_R\, \tilde{e}^-_R\qquad\quad
   [\gamma, Z\ \ \mbox{exchange}] \label{eq:se_L} \\
&& e^+_L\, e^-_R\ \ \to\ \ \tilde{e}^+_R\, \tilde{e}^-_R\qquad\quad
   [\gamma, Z, \tilde{\chi}^0\ \ \mbox{exchange}]
   \label{eq:se_R}
\end{eqnarray}
Though the signal process (\ref{eq:se_L}) is the analogue of smuon
pair production in $e^+e^-$ annihilation, we cannot anticipate that
in rival $J=1/2, 1,..$ processes $t$-channel exchanges do not
occur when the lepton number can flow from the initial to the final
state. Moreover, even electron polarization \cite{moortgat-pick} can
be realized only at a degree $< 1$ so that impurities from
the process are mixed in in any case. It is thus
plausible to evaluate $\tilde{e}^+_R\tilde{e}^-_R$ pair
production for unpolarized electron/positron beams.
[The general expression of the polarized cross section is given in the
Appendix.] This case exemplifies all the interesting characteristics.\s

After exploiting the conservation of the lepton current, the spinorial
parts of the matrix elements for $s$-channel $\gamma, Z$ exchange
and $t$-channel neutralino exchange are identical,
$\, \sim \, (\bar{v}_{e^+} \gamma_{\mu} u_{e^-})\,k^{\mu}$ with $k^{\mu}$
denoting the (spacelike) 4-momentum transfer. The $t$-channel
contribution can therefore be mapped into generalized charges,
introduced in analogy to Eqs.$\,$(\ref{eq:smuon_Q_L}) and (\ref{eq:smuon_Q_R}):
\begin{eqnarray}
&& Q_L = 1 + \left(s^2_W-1/2\right) c^{-2}_W D_Z(s)
   \label{eq:ql_ser} \\
&& Q_R = 1 + t^2_W D_Z(s)
          +\sum^4_{k=1} |N_{k1}|^2 c^{-2}_W D_{\tilde{\chi}^0_k} (t)
   \label{eq:qr_ser}
\end{eqnarray}
The indices $R$ and $L$ denote the electron helicities. The pole
part of the $\tilde{\chi}^0_k$ propagator, $k=1,..,4$ in the
Minimal Supersymmetric Standard Model (MSSM), is denoted in the center-of-mass
frame by
\begin{eqnarray}
D_{\tilde{\chi}^0_k}(t)\equiv \frac{s}{t-m^2_{\tilde{\chi}^0_k}}=
     \frac{-2}{\Delta_k-\beta_{\tilde{e}_R}\cos\theta}\quad
     \mbox{with}\quad
     \Delta_k = 1-2(m^2_{\tilde{e}_R}-m^2_{\tilde{\chi}^0_k})/s
\end{eqnarray}
while $N$ denotes the neutralino mixing matrix, see Ref.~\cite{choi_neutralino}.
Near the threshold, the $\tilde{e}^+_R\tilde{e}^-_R$ pair is produced
in a $P$-wave with amplitude $\sim\beta$. With rising energy however an
increasing number of orbital angular momenta is excited
and the propagator starts diverging in the forward direction
$\sim s/m^2_{\tilde{\chi}^0_k}$ [for $m_{\tilde{e}_R}> m_{\tilde{\chi}^0_k}$
after running through a maximum at
$\cos\theta \sim \Delta_k/\beta_{\tilde{e}_R}$].\s

The differential and total cross sections can be cast into the form
\begin{eqnarray}
  \frac{d\sigma}{d\cos\theta}[e^+e^-\to\tilde{e}^+_R\tilde{e}^-_R]
  &=& \frac{\pi\alpha^2}{8s}\beta^3\sin^2\theta\left[Q^2_L+Q^2_R\right]
  \label{eq:ser_differential_cross_section}
  \\
 \sigma[e^+e^-\to\tilde{e}^+_R\tilde{e}^-_R]
  &=& \frac{\pi\alpha^2}{8s}\beta^3
    \left[\langle Q^2_L\sin^2\theta\rangle
        + \langle Q^2_R\sin^2\theta\rangle\right]
  \label{eq:ser_total_cross_section}
\end{eqnarray}
with $\langle Q^2_{L,R}\sin^2\theta\rangle \equiv
\int d\cos\theta\, \sin^2\theta\, Q^2_{L,R}$.
Mass and energy dependence of the integrated charges can be adopted from
Ref.~\cite{RC2}:
\begin{eqnarray}
\langle Q^2_L\sin^2\theta\rangle
  &=& \frac{4}{3} \left[1+(s^2_W-1/2) c^{-2}_W D_Z(s)\right]^2  \\
\langle Q^2_R\sin^2\theta\rangle
  &=& \frac{4}{3}\, [1+t^2_W D_Z(s)]^2
   + 8 [1+t^2_W D_Z(s)] c^{-2}_W\sum^4_{j=1}|N_{j1}|^2 f^j\nonumber\\[-4mm]
  && \;\;\;\; + 8 c^{-4}_W \sum^4_{j,k=1}|N_{j1}N_{k1}|^2 h^{jk}
\end{eqnarray}
with the coefficients
\begin{eqnarray}
f^j=-\beta\Delta_j
   +\frac{\Delta^2_j-\beta^2}{2}\ln\frac{\Delta_j+\beta}{\Delta_j-\beta}\qquad
   \mbox{and}\qquad
h^{jk} =\left\{\begin{array}{ll}
   -2\beta +\Delta_j \ln\frac{\Delta_j+\beta}{\Delta_j-\beta}\ \  &\, j=k \\[2mm]
    (f^j-f^k)/(\Delta_j-\Delta_k)\ \ & \, j\neq k
    \end{array}\right.
\end{eqnarray}
\s

It follows that the production of $e$-type supersymmetric scalar particles
is characterized by the following two rules:
\begin{eqnarray}
  \#1  & \mbox{threshold excitation}  &\ \ \sim\ \ \beta^3
     \label{eq:rule_1sel}\\
  \#2  & \mbox{angular distribution}  &\ \ \sim\ \ \sin^2\theta \,\,
                                        \mathcal{G}(\cos\theta) \nonumber\\
{ }  & { }                          &\ \ \rightarrow\ \ \sin^2\theta\ \
                                      \mbox{near threshold}
\label{eq:rule_2sel}
\end{eqnarray}
Independent of energy, the angular distribution must behave
$\sim \sin^2\theta$ close to the forward and backward directions
where it must vanish by angular momentum conservation. While
this behavior may be masked in practice by the singularity in
$\mathcal{G}$ developing in the forward direction at high energies,
no such interference will arise in the backward direction. Since the
$\tilde{\chi}^0_k$ exchanges give rise to a $P$-wave near the threshold,
in the same way as $\gamma, Z$ exchange, a simple picture with
$\sigma\sim \beta^3$ and $d\sigma/d\cos\theta\sim\sin^2\theta$ emerges
at the threshold.\s

\begin{figure}[ht!]
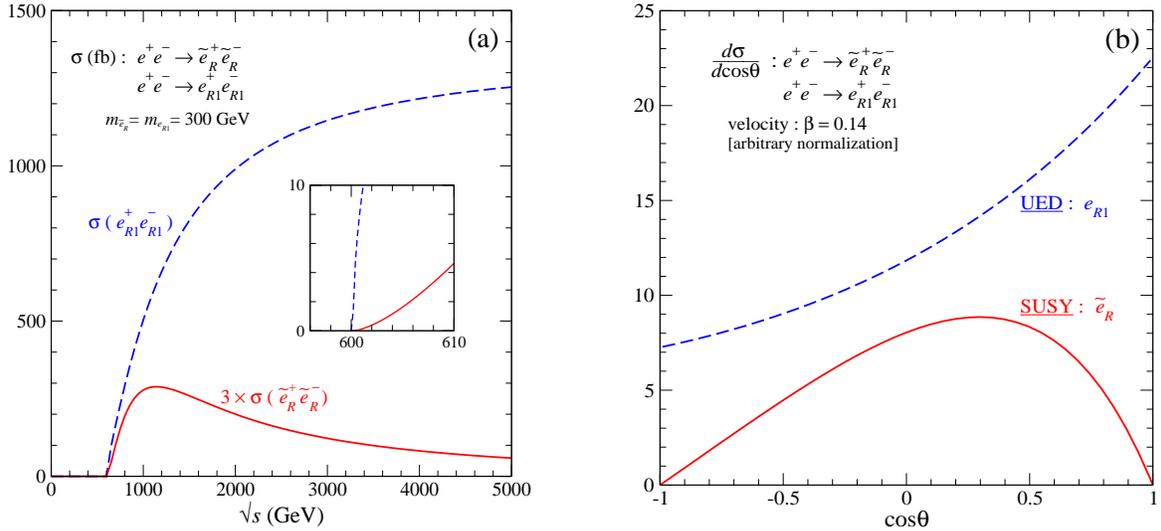

\begin{center}
\includegraphics[height=7.cm,width=7.cm,angle=0]{selectron_threshold.eps}
\hskip 1.2cm
\includegraphics[height=7.cm,width=7.cm,angle=0]{selectron_angle.eps}
\end{center}
\vskip -0.5cm \caption{\it (a) The threshold excitation for selectrons; and
             (b) the angular distribution in $e^+e^-\to\tilde{e}^+_R
             \tilde{e}^-_R$ for the SUSY parameters specified in the text.
             These results are compared with the production $e^+e^-\to e^+_{R1}
             e^-_{R1}$ of electronic Kaluza-Klein states in UED. The first KK
             mass $M_{K_1}$ is taken to be 295 GeV.}
\label{fig:selectron_threshold}
\end{figure}

Asymptotically the total cross section scales as
\begin{eqnarray}
\sigma\ \ \to\ \ \frac{\pi \alpha^2}{c_W^4}\,\, \frac{1}{s}\,
           \log {\frac{s}{m^2_{\tilde{e}_R}}}
           \quad  \mbox{for}\quad s\ \ \rightarrow\ \ \infty
\end{eqnarray}
as expected from the forward enhancement of the $t$-channel exchange. \s

These characteristics are displayed quantitatively in
Fig.$\,$\ref{fig:selectron_threshold}(a/b) for
the onset of the excitation curve and the angular distribution
close to threshold.\s\s

\subsection{KK Excited States \boldmath{$e^\pm_1$} in UED}

The analysis presented above repeats itself rather closely for the
KK excited states carrying electron lepton number; again we choose the
first KK excited R-type electrons $e^\pm_{R1}$ with vector couplings
to not only $\gamma$ and but also to $Z$ bosons as a representative example:
\begin{eqnarray}
e^+e^-\ \ \rightarrow\ \ e^+_{R1} e^-_{R1}
\end{eqnarray}
Analogously to Fig.$\,$\ref{fig:diagram1}(b), the $t$-channel exchange of the vector
and scalar KK excitations $B_1$ and $b_1$ [the supplement left with $B_1$
from the 5-dimensional vector] add to the standard $\gamma, Z$ exchanges
corresponding to Fig.$\,$\ref{fig:diagram1}(a).\s

Despite the complicated superposition of vector and scalar interactions,
Fierzing techniques allow us to cast all contributions into the
$s$-channel $\gamma_\mu \otimes \gamma_\mu$ form\footnote{The matrix
element $e^+e^-\to\mu^+_{R1}\mu^-_{R1}$ has earlier been defined
implicitly in the same way, the charges $Q_R$ and $Q_L$ to be identified with
$Q_R=Q_{RL}=Q_{RR}$ and $Q_L=Q_{LL}=Q_{LR}$ in this case.}
for the chiral $\alpha,\beta = L,R$ elements:
\begin{eqnarray}
{\cal M}[e^+e^-\to e^+_{R1}e^-_{R1}] = Q^1_{\alpha\beta}
   \left[\bar{v}(e^+)\gamma_\mu P_\alpha u(e^-)\right]
   \left[\bar{u}(e^-_{R1})\gamma^\mu P_\beta v(e^+_{R1})\right]
\end{eqnarray}
with the bilinear charges \cite{sehgal_zerwas}:
\begin{eqnarray}
\begin{array}{ll}
Q^1_{LL} = 1+\left(s^2_W-1/2\right) c^{-2}_W D_Z(s),\qquad\quad  &
Q^1_{LR} = Q^1_{LL}
          -\frac{1}{8}\frac{m^2_Z}{M^2_{K_1}+m^2_Z} t^2_W D_{b_1}(t)\\[2mm]
Q^1_{RL} = 1+t^2_W D_Z(s) + c_W^{-2} \frac{m^2_{e_{R1}}}{2 M^2_{K_1}}
                                            D_{B_1}(t),\qquad\quad &
Q^1_{RR} = 1+ t^2_W D_Z(s)+c^{-2}_W D_{B_1} (t)
\label{eq:bilinear_charge_1}
\end{array}
\end{eqnarray}
Apart from standard notations, $M_{K_1}$ denotes the first KK mass, and the
$t$-channel propagators are defined as $D_{B_1}(t)=D_{b_1}(t)
=s/[t-M^2_{K_1}]=-2/[1-2(m^2_{e_{R1}}-M^2_{K_1})/
s-\beta_{e_{R1}}\cos\theta]$.\s

After introducing the familiar quartic charges
\begin{eqnarray}
&& Q^1_1 = \frac{1}{4}\left[|Q^1_{RR}|^2+|Q^1_{LL}|^2
                           +|Q^1_{RL}|^2+|Q^1_{LR}|^2\right] \nonumber\\
&& Q^1_2 = \frac{1}{2}\real\left[Q^1_{RR}Q^{1*}_{RL}
                                +Q^1_{LL}Q^{1*}_{LR}\right] \nonumber\\
&& Q^1_3 = \frac{1}{4}\left[|Q^1_{RR}|^2+|Q^1_{LL}|^2
                           -|Q^1_{RL}|^2-|Q^1_{LR}|^2\right]
\label{eq:quartic_charge_1}
\end{eqnarray}
the differential cross section can be written in the compact form
\begin{eqnarray}
\frac{d\sigma}{d\cos\theta}\,[e^+e^-\to e^+_{R1} e^-_{R1}]
 = \frac{\pi\alpha^2}{2s}\beta
   \bigg[(1+\beta^2\cos^2\theta)Q^1_1+ (1-\beta^2) Q^1_2
          + 2\beta \cos\theta Q^1_3\bigg]
\label{eq:se_differential_cross_section}
\end{eqnarray}
from which the total cross section follows by integration over the polar
angle:
\begin{eqnarray}
\sigma[e^+e^-\to e^+_{R1} e^-_{R1}]
 = \frac{\pi\alpha^2}{2s}\beta
   \bigg[\langle Q^1_1\rangle +\beta^2\langle\cos^2\theta Q^1_1\rangle
         + (1-\beta^2) \langle Q^1_2\rangle
         + 2\beta \langle \cos\theta Q^1_3\rangle \bigg]
\label{eq:se_total_cross_section}
\end{eqnarray}
Both observables can serve as discriminants for Kaluza-Klein states against
supersymmetric selectrons. \s

By inspecting the cross sections in Eqs.$\,$(\ref{eq:se_total_cross_section}) and
(\ref{eq:se_differential_cross_section}) we can easily conclude, without studying
details, that
\begin{eqnarray}
\sigma[e^+e^-\to e^+_{R1}e^-_{R1}] &\sim& \beta\ \ \mbox{near threshold} \\[2mm]
\frac{d\sigma}{d\cos\theta}\, [e^+e^-\to e^+_{R1} e^-_{R1}]
  &\sim& (1+\beta^2\cos^2\theta)\, \mathcal{G}(\cos\theta) + \cdots \nonumber\\
  &\rightarrow& \mbox{flat in $\cos\theta$ near threshold}
\end{eqnarray}
as generally expected for fermion pair production near the threshold. As for
smuon pairs, these results contrast strongly to supersymmetric scalar
$\tilde{e}$ production. Most striking is the non-vanishing angular distribution
in the forward and backward directions. This is exemplified quantitatively in the
comparison of Fig.$\,$\ref{fig:selectron_threshold}(a/b).\s

Asymptotically however the total cross section, unlike the previous examples,
approaches a non-zero value
\begin{eqnarray}
\sigma\ \ \to\ \ \frac{\pi \alpha^2}{c_W^4} \, \frac{1}{M^2_{K_1}}
                       \quad  \mbox{for}\quad  s\ \ \rightarrow \ \ \infty
\end{eqnarray}
due to the enhancement in the forward direction, which is a remnant of the
Rutherford pole damped by the Yukawa mass cut-off
in the exchange of heavy particles. \s

\subsection{General Analysis}

Independent of the lepton number flow by additional $t$-channel exchange
mechanisms, the $s$-channel $\gamma, Z$ exchange in the production of charged
fermions of any spin $J=1/2, 3/2,\ldots$ will generate the pair in an
$S$-wave so that the cross section should rise at threshold $\sim \beta$
in contrast to the scalar $\beta^3$ particle production. The same $\gamma, Z$
exchange mechanism will generate a non-vanishing angular distribution in the
forward and backward directions $\neq\sin^2\theta$.\s

In contrast to the production of  muon-type pairs, the additional $t$-channel
exchanges in the production of $J=1,2,..$ integer spin electron-type
pairs will in general give rise to an $S$-wave component $\sim\beta$ in the onset
of the excitation curve. Since all spin $J>0$ particles in asymptotically
well-behaved field theories \cite{ferrara} will carry a non-vanishing magnetic
dipole moment, the angular distribution for both muon-type and electron-type
pairs will not vanish in forward/backward direction. This argument can be
supplemented by studying the polarizations $\sim d^J_{\lambda,\sigma}(\theta^*)$
in the decays of the two spin $J$ particles.\s

Thus in parallel to the smuon case, also for selectrons in
supersymmetric theories experimental paths can be designed for establishing
the scalar spin-0 character unambiguously.\s

\subsection{Simulation of \boldmath $\ee\to\ser^+\ser^-$}

The detection of scalar selectrons in the reaction
$\ee\to\ser^+\ser^-\to e^+\nt_1\,e^-\nt_1$
is again very clean. The event simulation, selection and analysis
proceed in complete analogy to smuon production,
described in the previous section,
just replacing the observable leptons by an $e^+e^-$ pair.\s

\begin{figure}[htb!]
\begin{center}
\includegraphics[width=17.cm,angle=0]{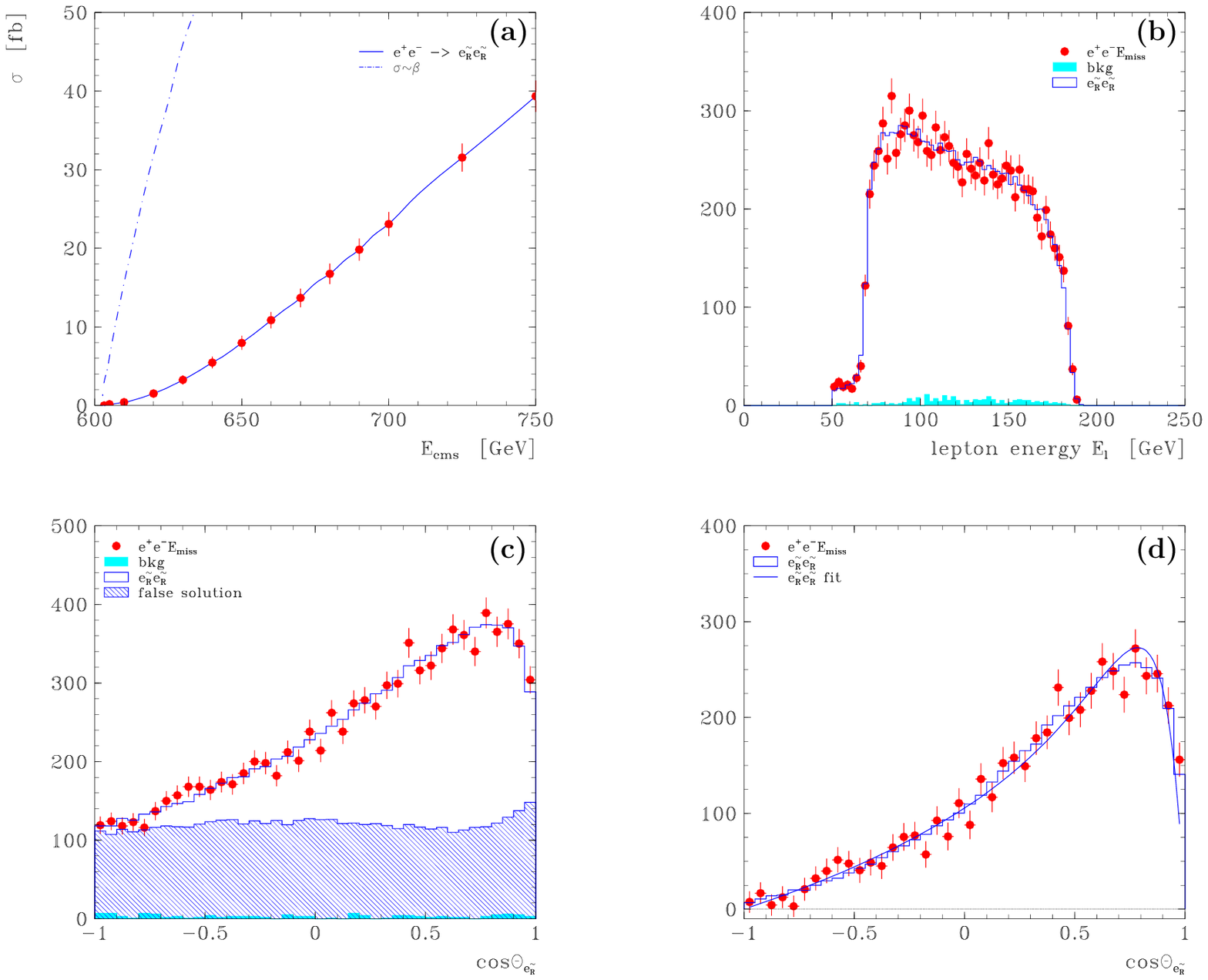}
\end{center}
\vskip -0.5cm
\caption{\it
  (a) The unpolarized cross section of $e^+e^-\to\ser^+\ser^-$ production
  close to threshold,
  including QED radiation, beamstrahlung and width effects;
  the statistical errors correspond to $\cL=10\,\fbi$ per point,
  the dash-dotted curve indicates a hypothetical dependence $\sigma\sim\beta$;
  (b) energy spectrum $E_e$ from $\ser^\pm\to e^\pm\nt_1$ decays;
  polar angle distribution $\cos\theta_\ser$
  (c) with and (d) without contribution of false solution.
  The simulation for the energy and polar angle distributions is based on
  polarized beams with $(\cP_{e^-},\cP_{e^+}) = (+0.8, -0.6)$
  at $\sqrt{s}=680\,\GeV$ and $\cL=200\,\fbi$.
  The smooth histograms represent high statistics expectations,
  the curve indicates a fit to the cross section
  (\ref{eq:ser_differential_cross_section}).}
\label{fig:figa2}
\end{figure}

The unpolarized cross section at threshold is displayed
in Fig.$\,$\ref{fig:figa2}(a).
The expected event rates are larger than for smuon production due to additional
$t$-channel neutralino exchange.
One observes a cross section typical for $P$-wave production
of spin-0 particles
with a dependence $\sigma_{\ser^+\ser^-}\sim\beta^3$,
as explained in Eq.$\,$(\ref{eq:rule_1sel}).
The excitation curve may be easily distinguished from a much faster rising
hypothetical $\sigma\sim\beta$ behavior, shown for comparison as well.\s

The study of the $\ser^+\ser^-$ production is performed close to threshold
in order to separate out, as well as possible, the factor $\sin^2\theta$ in
the polar-angle distribution, see Eq.$\,$(\ref{eq:rule_2sel}).
An energy of $\sqrt{s}=680\,\GeV$ appears to be a
good compromise between signal and background event rates.
An enhanced signal is obtained by choosing beam polarizations of
$(\cP_{e^-},\cP_{e^+}) = (+0.8, -0.6)$.
The cross section amounts to $47\,\fb$, and an integrated luminosity of
$\cL=200\,\fbi$ is assumed for the simulation.
The expected electron energy spectrum $E_e$ of the decays $\ser^\pm\to e^\pm\nt_1$
is shown in Fig.$\,$\ref{fig:figa2}(b). It exhibits a clear signature above a
negligible background from $W^+W^-$ and $\ser^\pm\sel^\mp$ production.\s

The angular distribution $\cos\theta_\ser$, shown in Fig.$\,$\ref{fig:figa2}(c),
is still peaked towards the forward direction, due to the remnant $t$-channel
$\tilde{\chi}^0$ contributions, above a fairly constant pedestal from the false
solutions. The spectrum after subtracting the ambiguity, displayed in
Fig.$\,$\ref{fig:figa2}(d), vanishes in the very forward and backward directions,
reflecting the overall $\sin^2\theta$ factor. A fit according to the differential
cross section formula (\ref{eq:ser_differential_cross_section}) yields a very
good description of the simulated data. As a by-product, the results of the fit
can be used to determine or cross-check the neutralino mixing parameters
$|N_{k1}|^2$ entering the expression (\ref{eq:qr_ser}) of the generalized
charge $Q_R$.\s

\begin{figure}[htb!]
\begin{center}
\includegraphics[width=\textwidth,height=7.5cm,angle=0]{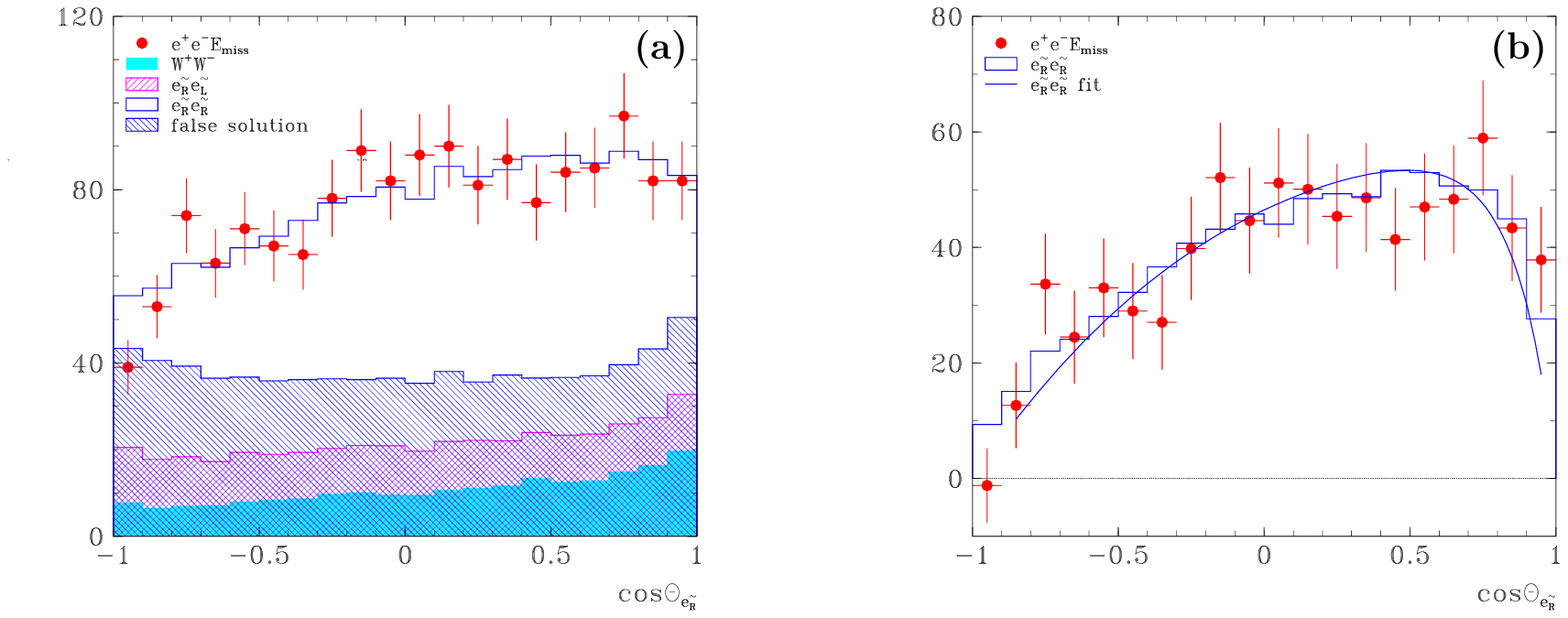}
\end{center}
\vskip -0.5cm
\caption{\it
  Polar angle distribution $\cos\theta_\ser$ of
  $e^+_Re^-_L\to\ser^+\ser^-$:
  (a) with contributions from background and false solution;
  (b) after subtraction of background and false solution.
  The simulation is based on polarized beams with
  $(\cP_{e^-},\cP_{e^+}) = (-0.8, +0.6)$
  at $\sqrt{s}=680\,\GeV$ and $\cL=300\,\fbi$.
  The smooth histograms represent high statistics expectations,
  the curve indicates a fit to the cross section
  (\ref{eq:ser_differential_cross_section})}
\label{fig:figa3}
\end{figure}

The $t$-channel neutralino exchange can be considerably reduced by choosing
opposite beam polarizations of $(\cP_{e^-},\cP_{e^+}) = (-0.8, +0.6)$.
These conditions however imply a much lower cross section of $\sigma=3.4\,\fb$
at the same energy $\sqrt{s}=680\,\GeV$
and a significantly larger background from $WW$ and $\ser\sel$ production.
The results of simulations assuming $\cL=300\,\fbi$ are displayed
in Fig.$\,$\ref{fig:figa3}.
The polar angle distribution is much flatter and shifted towards the central
region, approaching the expected $\sin^2\theta$ law for completely polarized beams.
A fit to the subtracted spectrum exhibits a skewed $\sin^2\theta$
distribution, reminiscent of small, residual $t$--channel contributions.\s

\section{CHARGINOS AND NEUTRALINOS}
\setcounter{equation}{0}

\subsection{Production Channels in \boldmath{$e^+e^-$} Collisions}

The prototypes of non-colored supersymmetric spin-1/2 fermions are the
charginos $\tilde{\chi}^\pm_1$ and the neutralinos $\tilde{\chi}^0_1$ and
$\tilde{\chi}^0_2$. They are produced in diagonal and mixed pairs in
$e^+e^-$ annihilation:
\begin{eqnarray}
&& e^+e^-\rightarrow \tilde{\chi}^+_1\tilde{\chi}^-_1  \hskip 2.9cm
 \mbox{with}\ \ \ \ \tilde{\chi}^\pm_1\rightarrow f\bar{f}'\tilde{\chi}^0_1 \\
&&
e^+e^-\rightarrow \tilde{\chi}^0_1\tilde{\chi}^0_2\ \ \mbox{and}\ \
                  \tilde{\chi}^0_2\tilde{\chi}^0_2  \hskip 1.1cm
 \mbox{with}\ \ \ \ \tilde{\chi}^0_2\rightarrow f\bar{f}\tilde{\chi}^0_1
\end{eqnarray}
Though a significant fraction of the decays is mediated potentially by
$\tilde{\tau}$ intermediate states as predicted in the reference
scenarios SPS1a/1a$'$ \cite{RS1,SPA,MG}, other decay modes can still play
a significant r$\hat{\rm o}$le due to large production cross sections,
in particular for diagonal pairs. The mixed neutralino production channel
$\tilde{\chi}^0_1\tilde{\chi}^0_2$
is easier to analyze in the threshold region when studying the
onset of the excitation curve, but the diagonal pair
$\tilde{\chi}^0_2\tilde{\chi}^0_2$ gives rise to a better textured
visible final state that allows the reconstruction of the flight
axis up to a 2-fold ambiguity, while the axis can be reconstructed for
mixed pairs only if $\tilde{\chi}^0_2$ cascades down through an intermediate
slepton \cite{cds}.\s\s

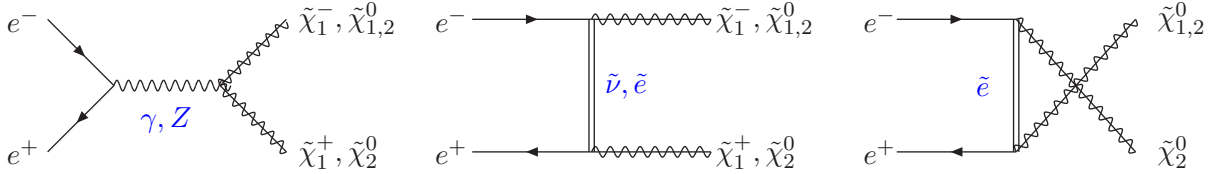
\begin{figure}[htb]
{
\begin{center}
\begin{picture}(450,70)(33,20)
\Text(47,75)[r]{$e^-$}
\ArrowLine(50,75)(75,50)
\ArrowLine(75,50)(50,25)
\Text(47,25)[r]{$e^+$}
\Photon(75,50)(119,50){2}{10}
\Text(95,37)[]{\color{blue} $\gamma,Z$}
\Line(115,50)(140,75)
\Photon(115,50)(140,74){2}{7}
\Text(145,75)[l]{$\tilde{\chi}^-_1,\tilde{\chi}^0_{1,2}$}
\Line(140,25)(115,50)
\Photon(140,25)(115,50){2}{7}
\Text(145,25)[l]{$\tilde{\chi}^+_1,\tilde{\chi}^0_2$}
\Text(209,75)[r]{$e^-$} \ArrowLine(210,75)(255,75)
\Text(209,25)[r]{$e^+$} \ArrowLine(255,25)(210,25)
\Line(254,75)(254,25) \Line(256,75)(256,25)
\Text(261,50)[l]{\color{blue} $\tilde{\nu}, \tilde{e}$}
\Line(255,75)(300,75) \Photon(255,75)(300,75){2}{8}
\Text(303,75)[l]{$\tilde{\chi}^-_1,\tilde{\chi}^0_{1,2}$}
\Line(300,25)(255,25) \Photon(300,25)(255,25){2}{8}
\Text(303,25)[l]{$\tilde{\chi}^+_1,\tilde{\chi}^0_2$}
\Text(365,75)[]{$e^-$}
\ArrowLine(370,75)(415,75)
\Text(365,25)[]{$e^+$}
\ArrowLine(415,25)(370,25)
\Line(414,75)(414,25)
\Line(416,75)(416,25)
\Text(404,50)[]{\color{blue} $\tilde{e}$}
\Line(415,75)(460,25)
\Photon(415,75)(460,25){2}{12}
\Text(470,75)[l]{$\tilde{\chi}^0_{1,2}$}
\Line(460,75)(415,25)
\Photon(460,75)(415,25){2}{12}
\Text(470,25)[l]{$\tilde{\chi}^0_2$}
\end{picture}
\end{center}
}
\caption{\it Two mechanisms contributing to the production of
   $\tilde{\chi}^+_1\tilde{\chi}^-_1$ and three to
   $\tilde{\chi}^0_1\tilde{\chi}^0_2$ and $\tilde{\chi}^0_2\tilde{\chi}^0_2$;
   $s$-channel $\gamma/Z$ exchanges and $t/u$-channel
   $\tilde{\nu}$ and $\tilde{e}$ exchanges, respectively.}
\label{fig:diagram2}
\end{figure}

Two mechanisms contribute to the production of $\tilde{\chi}^+_1\tilde{\chi}^-_1$
and three to $\tilde{\chi}^0_1\tilde{\chi}^0_2$ and
$\tilde{\chi}^0_2\tilde{\chi}^0_2$: $s$-channel $\gamma/Z$ exchanges
and $t/u$-channel $\tilde{\nu}$ and $\tilde{e}$ exchanges, respectively,
as illustrated in Fig.$\,$\ref{fig:diagram2}.\s

By using Fierzing techniques, both the $t/u$-channel diagrams can be mapped
onto the $s$-channel diagram, generating the bilinear charges $Q_{\alpha\beta}$
\cite{choi_chargino,choi_neutralino}:
\begin{eqnarray}
&& \begin{array}{lll}
        \mbox{\fbox{$\tilde{\chi}^+_1\tilde{\chi}^-_1$}}: &
        Q_{LL} = D_L- F_L \cos 2\phi_L &
        Q_{LR} = D'_L - F'_L \cos 2\phi_R \\
{ }   & Q_{RL} = D_R-F_R\cos 2\phi_L &
        Q_{RR} = D_R-F_R\cos 2\phi_R
        \end{array}
\label{eq:chargino_bilinear_charges}\\[2mm]
&& \begin{array}{ll}
        \mbox{\fbox{$\tilde{\chi}^0_i\,\,\tilde{\chi}^0_j$}}: &
        Q_{LL} = + 4 {\cal Z}_{ij} s^{-2}_{2W} (s^2_W-1/2) D_Z(s)
                 - {\cal G}_{Lij} D_{\tilde{e}_L}(u) \\
{ }   & Q_{LR} = - 4 {\cal Z}^*_{ij} s^{-2}_{2W} (s^2_W-1/2) D_Z(s)
                 + {\cal G}^*_{Lij} D_{\tilde{e}_L}(t) \\
{ }   & Q_{RL} = + {\cal Z}_{ij} c^{-2}_W D_Z(s)
                 + {\cal G}_{Rij} D_{\tilde{e}_R}(t) \\
{ }   & Q_{RR} = - {\cal Z}^*_{ij} c^{-2}_WD_Z(s)
                 - {\cal G}^*_{Rij} D_{\tilde{e}_R}(u)
        \end{array}
\label{eq:neutralino_bilinear_charges}
\end{eqnarray}
[in the usual notation $s^2_{2W} = \sin^2 2\theta_W$].
The normalized propagators in these charges read
\begin{eqnarray}
&& D_L = 1+  4 s^{-2}_{2W} (s^2_W-1/2)(s^2_W-3/4) D_Z(s)\qquad
   F_L =  4 s^{-2}_{2W} (s^2_W-1/2)D_Z(s)/4 \nonumber\\
&& D_R = 1+  c^{-2}_W (s^2_W-3/4) D_Z(s) \hskip 2.9cm
   F_R = c^{-2}_W D_Z(s)/4 \nonumber\\
&& D'_L = D_L + s^{-2}_W D_{\tilde{\nu}}(t)/4 \hskip 4.2cm
   F'_L = F_L - s^{-2}_W D_{\tilde{\nu}}(t)/4
\end{eqnarray}
and $\phi_{L,R}$ denote the mixing angles rotating the gaugino/higgsino current
to the chargino mass basis; the rotation angles are determined by the
SUSY Lagrangian parameters $M_2, \mu$ and $\tan\beta$ \cite{choi_chargino}.
The matrices ${\cal Z}$ and ${\cal G}_{L,R}$ are combinations of the mixing matrix
elements $N$ in the neutralino sector \cite{choi_neutralino}:
\begin{eqnarray}
&& {\cal Z}_{ij} = (N_{i3} N^*_{j3}-N_{i4}N^*_{j4})/2 \nonumber\\
&& {\cal G}_{Lij} = (N_{i2}c_W+N_{i1}s_W)(N^*_{j2}c_W
                        +N^*_{j1}s_W)/s^2_{2W}        \nonumber\\
&& {\cal G}_{Rij} = N_{i1}N^*_{j1}/c^2_W
\end{eqnarray}
They are derived from the Lagrangian parameters noted above and supplemented by
the U(1) gaugino parameter $M_1$ [in the MSSM].\s

Defining the quartic charges $Q_1, Q_2, Q_3$ in the same way as
Eq.$\,$(\ref{eq:quartic_charge_1}), the differential and total cross sections
can be written as
\begin{eqnarray}
\!\!\!\!\!\!\!\!\!\frac{d\sigma}{d\cos\theta} \!&=& \!\frac{\pi\alpha^2}{2s}\beta
 \left\{[1-(\mu^2_i-\mu^2_j)^2+\beta^2\cos^2\theta] Q_1
       +4\mu_i\mu_j Q_2
       +2\beta\cos\theta Q_3\right\}
\label{eq:ino_differential_cross_section} \\
\!\!\!\!\!\!\!\!\!\sigma \!&=& \!\frac{\pi\alpha^2f_s}{2s}\beta
 \left\{[1-(\mu^2_i-\mu^2_j)]\langle Q_1\rangle
        +\beta^2 \langle \cos^2\theta \,Q_1\rangle
        +4\mu_i\mu_j\langle Q_2\rangle
        +2\beta \langle \cos\theta \, Q_3 \rangle\right\}
\label{eq:ino_total_cross_section}
\end{eqnarray}
generically for any pair of masses with $\mu_i=m_i/\sqrt{s}$, and
$\beta^2=[1-(\mu_i-\mu_j)^2][1-(\mu_i+\mu_j)^2]$ coinciding with the
velocity squared for equal masses; $f_s = 1$ or $1/2$ denotes the
statistics factor for pairs of unequal and equal particles, respectively,
in the final state. \\[1.5mm]

\noindent
{\bf (a)} \underline{Charginos}:\\
\noindent
Near the threshold the cross section rises $\sim \beta$ since the charged
Dirac particles are generated in $S$-waves. For asymptotic energies the
cross section scales as
\begin{eqnarray}
\sigma\ \ \to\ \ \frac{4 \pi \alpha^2}{3 s}\,
   \left(1+\Delta_{\tilde{\chi}^\pm}\right)\quad
   \mbox{for}\quad  s\ \ \rightarrow\ \ \infty
\end{eqnarray}
complemented by the coefficient $\Delta_{\tilde{\chi}^\pm} ={\mathcal{O}}(1)$
including mixing matrix elements. \s

The angular distribution near the threshold is flat and need not vanish
as for scalar particles in the forward/backward direction. With rising
energy the sneutrino $t$-channel exchange excites an increasing number
of higher orbital angular momenta and thus modifies the familiar spin-$1/2$
asymptotic distribution to $\sim (1+\cos^2\theta)\, {\cal G}(\cos\theta)$.\s

The characteristics for supersymmetric chargino production can
therefore be summarized in the following points:
\begin{eqnarray}
& \#1\quad \mbox{threshold excitation}\ \ & \sim\ \ \beta
   \label{eq:chargino_characteristics_1}\\
& \#2\quad \mbox{angular distribution}\ \ & \sim\ \  (1+\beta^2\cos^2\theta)\,
                                                 \mathcal{G}(\cos\theta)
                                                 + \cdots\nonumber\\
&  &\rightarrow\ \ \mbox{isotropic near threshold}
   \label{eq:chargino_characteristics_2}
\end{eqnarray}
As will be argued later, these two characteristics can be mimicked
by higher half-integer spin states. Thus, the observation of the characteristics
(\ref{eq:chargino_characteristics_1}) and (\ref{eq:chargino_characteristics_2})
is necessary for chargino spin assignments in supersymmetric theories
but not sufficient. The production characteristics must be complemented
by \underline{decay characteristics} to determine the spin of charginos
unambiguously.\\[1.5mm]

\noindent
{\bf (b)} \underline{Neutralinos}:\\
\noindent
The Majorana nature of the neutralinos forbids the $S$-wave production
of the diagonal $\tilde{\chi}^0_2\tilde{\chi}^0_2$ pair at threshold
with equal spin components along the $e^+e^-$ beam axis as a consequence
of the Pauli principle.
This conclusion can also formally be drawn by observing that near threshold
the sum of the quartic charges is reduced to $Q_1 + Q_2 =
[|Q_{LL} + Q_{LR}|^2 + |Q_{RL} + Q_{RR}|^2$]/4 so that the final-state current
becomes purely vectorial, forbidden however for neutralino Majorana fields
which can only be coupled to axial-vector currents.
The $P$-wave production mode leads to the
onset of the excitation curve $\sim\beta^3$.
The angular distribution
follows the spin-1 rule for $\gamma/Z$ exchange \cite{Boud},
modified however by a spin-1 and spin-0 mixture from selectron $t/u$
exchanges. Inserting the quartic charges in
Eq.$\,$(\ref{eq:ino_differential_cross_section}), the
angular distribution is given near the threshold by
\begin{equation}
\frac{d \sigma_{thr}}{d \cos\theta} = \frac{\pi\alpha^2}{4s}\,\beta^3\,
 \sum_{k = L,R}
 \bigg[\!{\cal Z}^2_k (1 + \cos^2\theta)
  + 4 \tilde{g}_k(2 \tilde{g}_k - {\cal Z}_k) \sin^2{\theta}
  + 8 \delta_k \tilde{g}_k (\delta_k \tilde{g}_k + {\cal Z}_k) \cos^2\theta
  \, \bigg]
\label{eq:neutralino_thr_angle}
\end{equation}
where the coefficients are defined in terms of the matrices ${\cal Z}$ and
${\cal G}_{L,R}$ as
\begin{eqnarray}
&& {\cal Z}_R = {\cal Z} c^{-2}_W D_Z \hskip
3.15cm
  \tilde{g}_R = {\cal G}_R \, [1+\delta_R] \\
&& {\cal Z}_L = 4{\cal Z} s^{-2}_{2W} (1/2-s^2_W) D_Z \qquad\ \
  \tilde{g}_L = {\cal G}_L [1+\delta_L]
\end{eqnarray}
with $\delta_{L,R} =(m^2_{\tilde{\chi}^0_2} - m^2_{\tilde{e}_{L,R}})/
                    (m^2_{\tilde{\chi}^0_2} + m^2_{\tilde{e}_{L,R}})$.
For large selectron masses the $s$-channel $Z$ contributions are dominant and
the angular distribution is reduced
to $(1+\cos^2\theta)$, characteristic for Majorana fermion pair production.
However, if for particle masses of the same size,
the $t-$ and $u-$channel selectron contributions are
dominant, the angular distribution
is in general a mixture of $\sin^2\theta$ and $\cos^2\theta$ terms with
coefficients varying with the particle masses.
Above the threshold, higher orbital angular momenta are excited by the selectron
exchange mechanisms, not altering however the asymptotic behavior
\begin{eqnarray}
\sigma\ \ \to\ \ \frac{4 \pi \alpha^2}{3 s}\, \Delta_{\tilde{\chi}^0}
    \quad  \mbox{for}\quad  s\ \ \rightarrow\ \ \infty
\end{eqnarray}
with the coefficient $\Delta_{\tilde{\chi}^0} = {\mathcal{O}}(1)$. \s

Also mixed $\tilde{\chi}^0_1 \tilde{\chi}^0_2$ pairs will be produced
near the threshold in a $P$-wave if their CP parities, $\pm i$, are equal.
If they are different however $S$-wave production is possible and the cross
section rises $\sim\beta$. In theories with CP violation $S$-wave production is
predicted in general \cite{choi_neutralino,Choi:2003hm}.\s

These observations are summarized in the following rules:
\begin{eqnarray}
&& \#1\quad \mbox{threshold excitation}\ \ \sim\ \ \beta^3\ \ \mbox{for}\ \
                     \tilde{\chi}^0_2\tilde{\chi}^0_2
                      \label{eq:neutralino_charactersistics_1}\\
&& { }\hskip 4.23cm \ \ \sim\ \ \beta^3 / \beta\ \ \mbox{for}\ \
                     \tilde{\chi}^0_1\tilde{\chi}^0_2\
                     \ [\,\mbox{ident./diff. Majorana phases}\,]\ \
                     \nonumber\\
&& \#2\quad \mbox{angular distribution near threshold}\ \
                     \sim\ \ \cos^2/\sin^2\theta \; \mbox{mix}\
   \label{eq:neutralino_characteristics_2}
\end{eqnarray}

\begin{figure}[ht!]
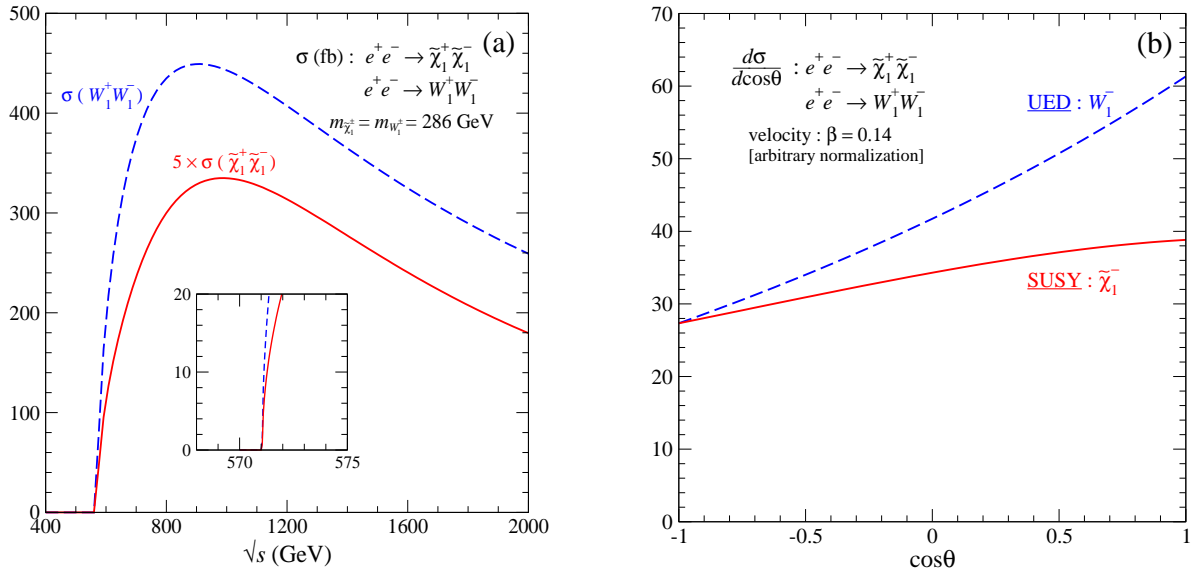

\begin{center}
\includegraphics[height=7.5cm,width=7.2cm,angle=0]{chargino_threshold.eps}
\hskip 1.2cm
\includegraphics[height=7.5cm,width=7.2cm,angle=0]{chargino_angle.eps}
\end{center}
\vskip -0.5cm \caption{\it (a) The threshold excitation for charginos
                 and (b) the angular distribution in the process
                 $e^+e^-\to\tilde{\chi}^+_1 \tilde{\chi}^-_1$
                 for the SUSY parameters specified in the text;
                 both compared with $W^+_1 W^-_1$ pair production
                 in UED.}
\label{fig:chargino_threshold}
\end{figure}

\begin{figure}[ht!]
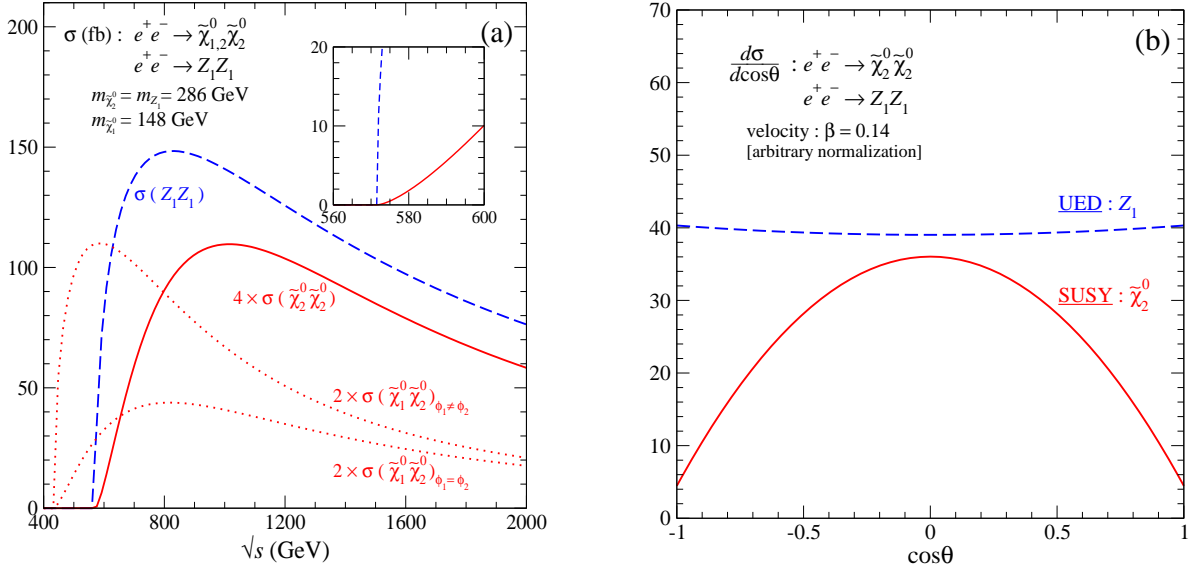

\begin{center}
\includegraphics[height=7.5cm,width=7.2cm,angle=0]{neutralino_threshold.eps}
\hskip 1.2cm
\includegraphics[height=7.5cm,width=7.2cm,angle=0]{neutralino_angle.eps}
\end{center}
\vskip -0.2cm
\caption{\it (a) The threshold excitation for
              neutralinos and (b) the angular distribution in the processes
              $e^+e^- \to \tilde{\chi}^0_1\tilde{\chi}^0_2$ and $
             \tilde{\chi}^0_2\tilde{\chi}^0_2$
             for the SUSY parameters specified in the text;
             both compared with $Z_1 Z_1$ pair production
             in UED.}
\label{fig:neutralino_threshold}
\end{figure}

These points are illustrated for charginos and neutralinos in
Figs.$\,$\ref{fig:chargino_threshold} and \ref{fig:neutralino_threshold},
respectively. The parameter set introduced earlier, gives rise to the
chargino mass $m_{\tilde{\chi}^\pm_1} =$ 286 GeV and the neutralino
masses $m_{\tilde{\chi}^0_{1/2}} = 148/286$ GeV.
The residual linear $\beta$-dependence, $\beta = 0.14$, generates
a slight increase of the angular distribution with $\cos\theta$. For neutralinos
the chosen parameter set leads to a dominant $\sin^2\theta$ component in the
angular distribution, supplemented however by small additional contributions,
{\it cf.} Eq.$\,$(\ref{eq:neutralino_thr_angle}), which
render the distributions non-vanishing at the very edges of the forward and
backward directions.
These results will be confronted with phenomena
in UED and a general analysis in the next subsections.\s

%

\subsection{KK Excited States \boldmath{$W^\pm_1$} and
  \boldmath{$Z_1/\gamma_1$} in UED}

The counterpart of $\tilde{\chi}^\pm_1$ and $\tilde{\chi}^0_2$ in UED are
the KK excitations $W^\pm_1$ and $W^3_1=Z_1$, while $\tilde{\chi}^0_1$ and
$B_1=\gamma_1$ are the stable particles of the two theories with minimal
mass\footnote{Electroweak mixing at the KK level is neglected as before.}
to which all other particles cascade down.\s\s

\begin{figure}[!h]
{
\begin{center}
\begin{picture}(220,180)(50,20)
\Text(25,175)[r]{$e^-$}
\ArrowLine(30,175)(55,150)
\ArrowLine(55,150)(30,125)
\Text(25,125)[r]{$e^+$}
\Photon(55,150)(99,150){2}{10}
\Text(75,137)[]{\color{blue} $\gamma,\,Z$}
\Photon(95,150)(120,175){2}{10}
\Text(135,175)[]{$W^-_1$}
\Photon(120,125)(95,150){2}{10}
\Text(135,125)[]{$W^+_1$}
\Text(200,175)[]{$e^-$}
\ArrowLine(210,175)(255,175)
\Text(200,125)[]{$e^+$}
\ArrowLine(255,125)(210,125)
\ArrowLine(254,175)(254,125)
\Text(244,150)[]{\color{blue} $\nu_1$}
\Photon(255,175)(300,175){2}{10}
\Text(310,175)[l]{$W^-_1$}
\Photon(300,125)(255,125){2}{10}
\Text(310,125)[l]{$W^+_1$}
\Text(25,75)[r]{$e^-$}
\ArrowLine(30,75)(75,75)
\ArrowLine(75,25)(30,25)
\Text(25,25)[r]{$e^+$}
\ArrowLine(75,75)(75,25)
\Text(55,50)[l]{\color{blue} $e_{L1}$}
\Photon(75,75)(120,75){2}{10}
\Text(147,75)[]{$W^3_1=Z_1$}
\Photon(75,25)(120,25){2}{10}
\Text(147,25)[]{$W^3_1=Z_1$}
\Text(200,75)[]{$e^-$}
\ArrowLine(210,75)(255,75)
\Text(200,25)[]{$e^+$}
\ArrowLine(255,25)(210,25)
\ArrowLine(254,75)(254,25)
\Text(242,50)[]{\color{blue} $e_{L1}$}
\Photon(255,75)(300,25){2}{10}
\Text(305,75)[l]{$W^3_1=Z_1$}
\Photon(300,75)(255,25){2}{10}
\Text(305,25)[l]{$W^3_1=Z_1$}
\end{picture}
\end{center}
}
\vskip 0.2cm
\caption{\it Two mechanisms contributing to the production of
   $W^+_1W^-_1$ and two to $Z_1Z_1$; $s$-channel $\gamma/Z$ exchanges,
   $t$-channel $\nu_1$ and $t/u$-channel $e_{L1}$ exchanges, respectively.}
\label{fig:diagram3}
\end{figure}
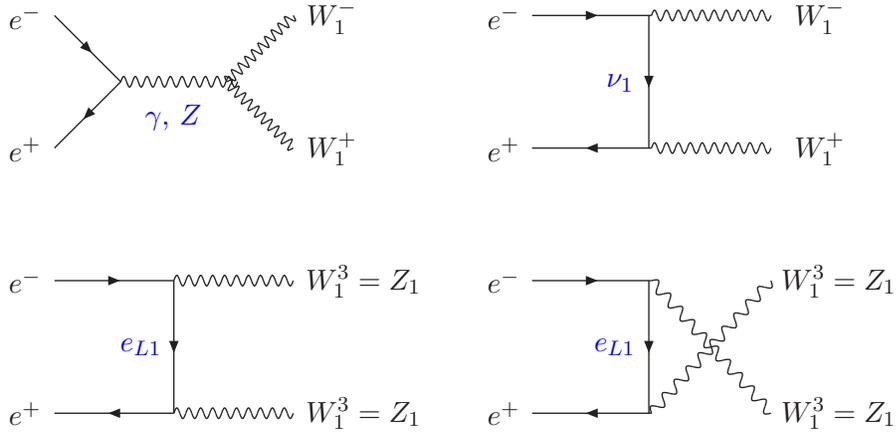

The cross sections for the processes
\begin{eqnarray}
&& e^+e^-\ \ \rightarrow\ \ W^+_1 W^-_1 \\
&& e^+e^-\ \ \rightarrow\ \ \, Z_1 Z_1
\end{eqnarray}
are closely related to the corresponding SM processes $e^+e^-\to W^+W^-$
and $ZZ$, {\it cf.} Fig.$\,$\ref{fig:diagram3}. Standard SM couplings are
attached to the currents, and the exchanged neutrino and electron must be
substituted by the heavy KK excitations. In the limit in which the masses of
the $t/u$-exchange leptons are neglected, the cross sections approach the SM
form of Refs.~\cite{AB,Hag}. \s

The differential and total cross sections for $W^+_1W^-_1$
can be expressed by the generalized charges
\begin{eqnarray}
Q_L = 1-(s^2_W-1/2) s^{-2}_W D_Z(s)
\quad\ \ \mbox{and}\quad\ \ Q_R = 1- D_Z(s)
\end{eqnarray}
In this notation they can be written as
\begin{eqnarray}
\frac{d\sigma}{d\cos\theta}\! &=&\! \frac{\pi\alpha^2}{8s}\beta
    \left[s^{-4}_WF_1(s,\theta)+\frac{1}{2}(Q^2_L+Q^2_R)\,F_2(s,\theta)
         - Q_{L} s_W^{-2} F_3(s,\theta)\right]
\label{eq:differential_cross_section_w_pm_1} \\
\sigma \!&=&\! \frac{\pi\alpha^2}{8s}\beta
   \left[s^{-4}_W \, \bar{\sigma}_1(s)+\frac{1}{2}(Q^2_L+Q^2_R)\,\bar{\sigma}_2(s)
         - Q_{L} s_W^{-2} \, \bar{\sigma}_3(s)\right]
\label{eq:total_cross_section_w_pm_1}
\end{eqnarray}
The angular functions $F(s,\theta)$ and the
energy-dependent coefficients $\bar{\sigma} (s)$ are given by
\begin{eqnarray}
&& F_1(s,\theta) = \frac{\gamma^2(1+\beta^2-2\beta\cos\theta)^2
     \, (4+\gamma^2\beta^2\sin^2\theta)+4\beta^2\sin^2\theta}{
    4(\Delta - \beta\cos\theta)^2} \\[2mm]
&& F_2(s,\theta) = 2\beta^2\left[4\gamma^2(4+\gamma^2\beta^2\sin^2\theta)
    + 3\sin^2\theta\right]\\[2mm]
&& F_3(s,\theta) = \frac{2\gamma^2\beta \, [2\beta-(1+\beta^2)\cos\theta]
     (4+\gamma^2\beta^2\sin^2\theta)+3\beta^2\sin^2\theta}{
     (\Delta -\beta\cos\theta)}
\end{eqnarray}
with $\gamma=\sqrt{s}/2m_{W^\pm_1}$, $\Delta = 1-2(m^2_{W^\pm_1}-m^2_{\nu_1})/s$,
and
\begin{eqnarray}
\bar{\sigma}_1(s) &=& 4(\gamma^4+11\gamma^2-3)/3-(2\gamma^2-1)\eta^2-2\eta^4
    -2\left[\gamma^{-2}+2(2-\gamma^{-2})\eta^2\right]/(\Delta^2-\beta^2)
    \nonumber\\
    && +\left[4\Delta-(7+\Delta)\eta^2+(1+2\Delta)\eta^4\right]
       L(\bar{\beta})/\Delta\\
\bar{\sigma}_2(s) &=& 8\beta^2
    (4\gamma^4+20\gamma^2+3)/3\\
\bar{\sigma}_3(s) &=& 6\Delta +8(2\gamma^4+9\gamma^2-5)/3
    -[\gamma^{-2}+(2-\gamma^{-2})\eta^2]\,(2\gamma^2-1+\eta^2)
    \nonumber\\
   && -\left\{6(\Delta^2-\beta^2)
      + \left[\gamma^{-2}+(2-\gamma^{-2})\eta^2\right]
        \left[7+\Delta-(1+\Delta)\eta^2\right]\right\}\, L(\bar{\beta})/\Delta
\end{eqnarray}
with $\eta=m_{\nu_1}/m_{W^\pm_1}$, $\bar{\beta}=\beta/\Delta$ and $L(x)=
(1/2x)\log[(1+x)/(1-x)]\rightarrow 1$ as $x\rightarrow 0$. Although
each of the individual coefficients $\bar{\sigma}_i(s)$ grows as $s$,
unitarity cancellations reduce the sum of all contributions
to the expected scaling behavior of the cross section \cite{AB,Hag}:
\begin{eqnarray}
\sigma[e^+e^-\to W^+_1 W^-_1] \ \ \rightarrow\ \ \frac{\pi\alpha^2}{2s^4_W}\,
\frac{1}{s}\,\log \frac{s}{m^2_{W^\pm_1}}\quad  \mbox{for}\quad
 s\rightarrow \infty
\end{eqnarray}
The logarithmic term is generated by the KK neutrino exchange mechanism. \s

For $Z_1Z_1$ production the differential and total cross sections read
\begin{eqnarray}
&&\mbox{ }\hskip -1.7cm  \frac{d\sigma}{d\cos\theta}
   = \frac{\pi\alpha^2}{16 s^4_W s}\beta
 \left\{ \frac{2-\beta^2(1+\cos^2\theta)}{
          \Delta^2-\beta^2\cos^2\theta}
          +\frac{2\beta^4\sin^2\theta\cos^2\theta}{
        (\Delta^2-\beta^2\cos^2\theta)^2}
  +\eta^4\,
  \frac{\beta^2\cos^2\theta \left[4(1-\beta^2)+\beta^2\sin^2\theta\right]}{
        2(\Delta^2-\beta^2\cos^2\theta)^2}
         \right\} \\[0mm]
 &&\mbox{ }\hskip -0.8cm\sigma = \frac{\pi\alpha^2}{32 s^4_W s}\beta\,
 \bigg\{\, 8(1+\bar{\gamma}^{-2}) L\left(\bar{\beta}\right)-8
 +\eta^4
     \left[(3-\bar{\beta}^2-4\bar{\gamma}^{-2}) L\left(\bar{\beta}\right)
         -3+4\bar{\gamma}^{-2}/(1-\bar{\beta}^2)\right] \bigg\}
\end{eqnarray}
with $\eta=m_{e_{L1}}/m_{Z_1}$, $\Delta=1-2(m^2_{Z_1}-m^2_{e_{L1}})/s$ and
$\bar{\gamma}=\gamma \Delta$. For asymptotically large energies, the standard
behavior
\begin{eqnarray}
\sigma[e^+e^-\to Z_1 Z_1] \ \ \to\ \ \frac{\pi\alpha^2}{8s^4_W}\,
\frac{1}{s}\, \log \frac{s}{m^2_{Z_1}}\quad  \mbox{for}\quad
 s\ \ \rightarrow\ \ \infty
\end{eqnarray}
is predicted for the total cross section. \s

Near the thresholds the total cross sections rise as
\begin{eqnarray}
\sigma[e^+e^-\to W^+_1W^-_1, Z_1 Z_1]\ \ \sim\ \ \beta
\end{eqnarray}
while the angular distributions
\begin{eqnarray}
 \frac{1}{\sigma}\,\frac{d\sigma}{d\cos\theta}[W^+_1W^-_1]
  \ \ \simeq \ \ \frac{1}{2}
         + O(\beta) \cos\theta\quad\, \mbox{and}\,\quad
\frac{1}{\sigma}\,\frac{d\sigma}{d\cos\theta}[Z_1Z_1]
  \ \ \simeq \ \
          \frac{1}{2}
         + O(\beta) \cos^2\theta
\end{eqnarray}
are essentially flat in the threshold region.
The flat behavior is modified however linearly in $\beta$ above the threshold
as evident from Fig.$\,$\ref{fig:chargino_threshold}(b). \s

Comparing the predictions for the spin-1 KK excitations of the weak gauge
bosons with the spin-1/2 charginos and neutralinos, we arrive at
a mixed picture, {\it cf.} Figs.$\,$\ref{fig:chargino_threshold} and
\ref{fig:neutralino_threshold}.
In the chargino sector the onset of the excitation
curves does not discriminate one from the other. However, due to
the Majorana nature of the neutralinos, the onset for
$\tilde{\chi}^0_2\tilde{\chi}^0_2\sim\beta^3$ is different from $Z_1Z_1
\sim\beta$.\s

Final state analyses are necessary to discriminate charginos from KK $W^\pm_1$
bosons. Due to the vectorial/axial-vectorial couplings, in both theories, the
electron-positron pair annihilates in a spin-1 state polarized parallel
to the beam axis. Angular momentum conservation then demands the same
polarization state for the charginos which are coupled in an $S$-wave.
Choosing longitudinally polarized electron beams with a degree close
to one \cite{moortgat-pick}, the decay angular distribution is dictated by
\begin{eqnarray}
{\cal D}[\tilde{\chi}^-_1\to(f\bar{f}')\tilde{\chi}^0_1]\ \ \sim\ \
d^{1/2}_{\lambda \sigma}(\theta^*)\ \ \sim\ \
\cos(\theta^*/2)\ \ \mbox{or}\ \
\sin(\theta^*/2)
\end{eqnarray}
[depending on whether the initial $\tilde{\chi}^-_1$ helicity $\lambda$
and the difference $\sigma=\sigma(\tilde{\chi}^0_1)-\sigma(f\bar{f}')$ of
final-state helicities are of equal or opposite sign]. This can be contrasted to
the polarization of the $W^-_1$ which must be either 1 or 0, so that, in the same
notation as before,
\begin{eqnarray}
{\cal D}[W^-_1\to (f\bar{f}') \gamma_1]\ \ \sim\ \
 d^1_{\lambda\sigma}(\theta^*)\ \ \sim\ \
(1\pm\cos\theta^*)\ \ \mbox{or}\ \ \cos\theta^*\ \ \mbox{or} \ \
\sin\theta^*
\end{eqnarray}
[depending on whether $|\lambda|=|\sigma|=1, 0$ or otherwise]
with quite a different Wigner $d$ function compared to the
supersymmetric signal. Thus the final
state analysis provides a clear discrimination.\s

In conclusion. The supersymmetric chargino/neutralino sector can be discriminated
in spin analyses from the KK excited weak-boson sector in
theories of universal extra space dimensions, but final state analyses of the
decaying particles are required.\s

\subsection{General Analysis}
\label{sec:4.3}

\noindent
{\bf (a)} \underline{Charginos}: $S$-wave production of chargino pairs
gives rise to the $\beta$ onset of the excitation curve near the threshold.
This behavior is expected for all charged half-integer spin Dirac
particles. In parallel, the angular distribution in the production process
does not discriminate the particles. Bosons with spin $\geq 1$ also
follow the $S$-wave pattern if they are produced pairwise through
$t$- and/or $u$-channel exchanges.\s

Quite generally, the (polarized) electron/positron pair either annihilates in
a polarized spin-1 state for vector currents, as exemplified
above.\footnote{Scalar and tensor couplings, which would correspond to spin-0
states, vanish in the limit of zero electron mass due to electric chirality
conservation.}
This gives rise to polarization
effects  $d^J_\lambda(\theta^*)$ in the $F^J$ decays and to correlations
of the form
\begin{eqnarray}
d^J_{\lambda\pm 1,\sigma}(\theta^*_1)\, d^J_{\lambda \sigma'} (\theta^*_2)
   \qquad  \mbox{and/or}\qquad
d^J_{\lambda \sigma}(\theta^*_1)\, d^J_{\lambda \sigma'}(\theta^*_2)
\end{eqnarray}
between the angular distributions of the decay products of the particle pair
$F^J\bar{F}^J$ which is generated in an $S$-wave near threshold. The characteristic
dependence of the $d$ functions on $J$ can be exploited to determine the spin.
[Details will be presented in the subsequent experimental subsection.]\s

\noindent
{\bf (b)} \underline{Neutralinos}: For clarity we focus on the production
process (\ref{eq:general_process}) for equal-type particle-antiparticle
$F^J\neq \bar{F}^J$ and particle-particle $F^J=\bar{F}^J$ pairs. As
argued before, $S$-wave production
is expected in general if the neutral fermions $F^J$ and $\bar{F}^J$ are
different from each other; it gives rise to the $\beta$ dependence
of the cross section near threshold as opposed to the $\beta^3$ production
law of the Majorana particle $\tilde{\chi}^0_2$. \s

It has been shown quite generally in Ref.$\,$\cite{Boud} that Majorana pairs
$F^JF^J$ are always produced in $P$ waves near threshold, with a
$(1 + \cos^2\theta)$
angular distribution for spin-1 $\gamma/Z$ exchange. If $t/u$-channel exchanges
are switched on, $S$-wave production of Majorana fermion pairs remains suppressed for
all interactions conserving electron-chirality. While the rise of the excitation
curve $\sim \beta^3$ does not change, the angular distribution is modified however to
a mix of $\cos^2\theta$ and $\sin^2\theta$ terms. \s

Thus the spin of charginos and neutralinos cannot be discriminated unambiguously
unless the standard correlation tests involving the chargino/neutralino decays
with reasonable polarization analysis power are performed.
The analysis of polarization effects in $F^J$ decays, eventually supplemented
by correlation effects in double
$F^J\bar{F}^J$ and $F^JF^J$ decays, in the way discussed above,
will lead to the unambiguous spin assignment $J=1/2$ of the
$\tilde{\chi}^\pm_1$ chargino and the $\tilde{\chi}^0_2$ neutralino.\s

\subsection{Simulation of \boldmath $\ee\to\cx^+_1\cx^-_1$
  and $\ee\to\nt_2\nt_2$}

Chargino production and detection proceeds via
$\ee\to\cx^+_1\cx^-_1\to W^+\nt_1\;W^-\nt_1$
with a branching ratio $\cB(\cx^\pm_1\to W^\pm\nt_1)=1$
in the reference point considered.
Distinct experimental signatures are either purely hadronic decays
$WW\to q\bar q'\; \bar q q'\to 4\, {\rm jets}$ or mixed hadronic and leptonic
decays $WW\to q\bar q'\;\ell\nu \to 2~{\rm jets}+ 1~{\rm lepton}$.
For a complete reconstruction of the kinematics, including
production and decay angles, only the 4-jet final state can be used.
However, information on the individual $W^\pm$ charge is heavily spoiled by
large fluctuations during the fragmentation process which may lead
to track losses and/or wrong track assignments to the parent particle.
Only a folded angle $|\cos\theta|$ can be obtained. In contrast,
the electric charge of individual $W^\pm$ can be identified in the mixed
hadronic and leptonic decays, $W^\pm W^\mp\to q\bar{q}' \ell^\mp \nu$.\s

A potential background source is neutralino production $\ee\to\nt_2\nt_2$
with subsequent decays $\nt_2\to Z\nt_1~(\cB=0.13)$ and
$\nt_2\to h\nt_1~(\cB=0.87)$. The hadronic decays of $Z$ and $h$ provide an
event topology and kinematics very similar to chargino production.
A distinction may be possible on the basis of excellent di-jet mass resolution
as anticipated in the design of future ILC detectors~\cite{ldc}.
The goal is to achieve an efficient separation of hadronic $W$ and $Z$ decays,
which also implies a reliable identification of the heavier Higgs decays.\s

\begin{figure}[htb!]
\begin{center}
\includegraphics[width=\textwidth,angle=0]{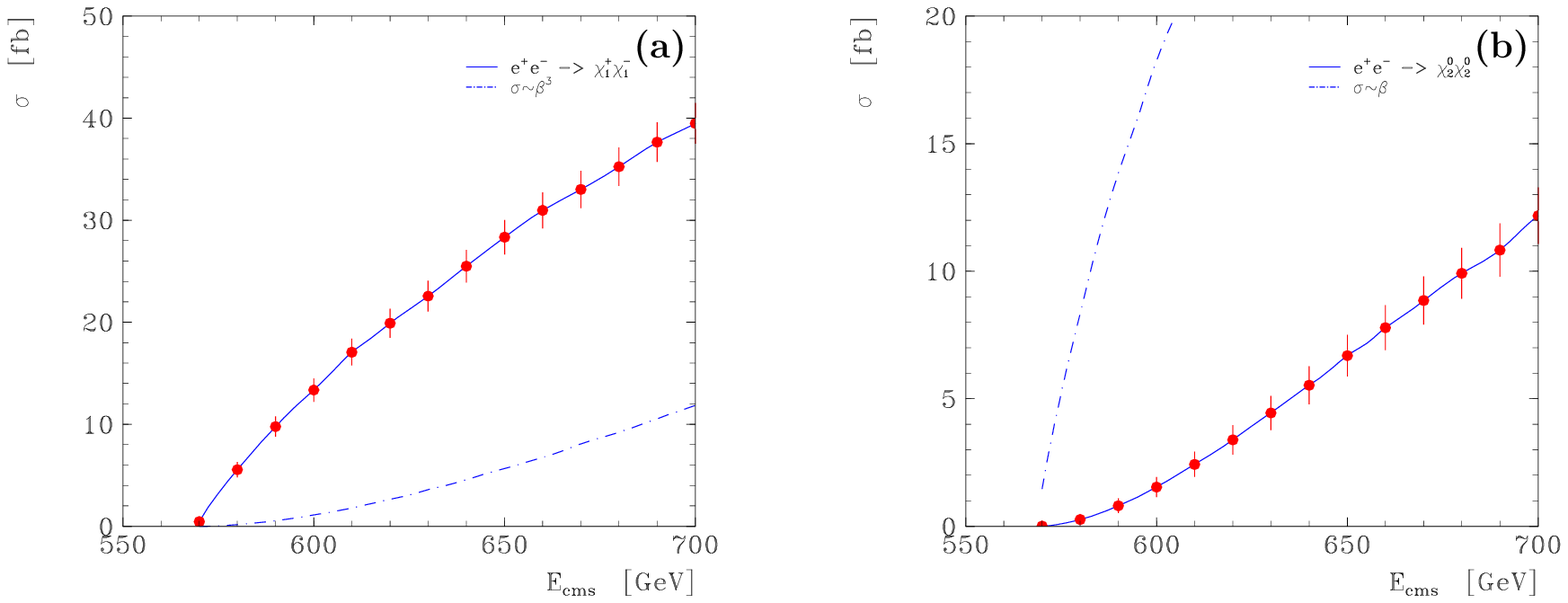}
\end{center}
\vskip -0.5cm
\caption{\it
  Cross sections of
  (a) $e^+e^-\to\cx^+_1\cx^-_1$ production  and
  (b) $\ee\to\nt_2\nt_2$ production close to threshold,
  including QED radiation, beamstrahlung and width effects.
  The statistical errors correspond to $\cL=10\,\fbi$ per point,
  The dash-dotted curves indicate hypothetical cross section dependencies
  $\sigma_{\cx^+_1\cx^-_1}\sim\beta^3$ and
  $\sigma_{\cx^0_2\cx^0_2}\sim\beta$
  for charginos and neutralinos, respectively.}
\label{fig:figa4}
\end{figure}

The cross sections for $\cx^+_1\cx^-_1$ and $\cx^0_2\cx^0_2$ production
as a function of energy are shown in Fig.$\,$\ref{fig:figa4}(a) and
Fig.$\,$\ref{fig:figa4}(b), respectively.
Since the masses are almost degenerate, the threshold energies of both reactions
are very close and practically coincide.
However, the chargino cross section rises much faster with
$\sigma\sim\beta$ compared with the slow onset of the neutralino excitation
curve $\sigma\sim\beta^3$.
It is obvious from the threshold curves that the different (opposite)
$\beta$ dependence for the two reactions
can be easily ruled out.\s

As pointed out in the previous section, polarization
effects in $\tilde{\chi}^\pm_1$ and $\tilde{\chi}^0_2$ decays must be exploited
to determine the spin $J=1/2$ of the $\tilde{\chi}^\pm_1$ chargino and
$\tilde{\chi}^0_2$ neutralino unambiguously. \s

\noindent
{\bf (a)} \underline{Charginos $\tilde{\chi}^\pm_1$}:\\
The charginos in the production process $e^+e^-\to\tilde{\chi}^-_1
\tilde{\chi}^+_1$ are polarized and even the polarization averaged over the
production angle $\theta$ is in general non-zero (also for unpolarized beams).
The cosine of the decay angle $\theta^*_{W^\pm}$ between the $W^\pm$ momentum
direction in the chargino rest frame and the $\tilde{\chi}^\pm_1$ momentum direction
in the laboratory frame, identical with the spin quantization axis, can be
determined by measuring the $W^\pm$ energy in the hadronic $W^\pm$ decay
$W^\pm\to q \bar{q}'$:
\begin{eqnarray}
E_{W^\pm} = \gamma \left( E^*_{W^\pm}
                        +\beta\, p^*_{W^\pm}\cos\theta^*_{W^\pm}\right)
\label{eq:energy_angle_relation}
\end{eqnarray}
where $\gamma=\sqrt{s}/2m_{\tilde{\chi}^\pm_1}$ and $\beta=(1-
4 m^2_{\tilde{\chi}^\pm_1}/s)^{1/2}$. The $W^\pm$-boson energy and momentum
in the chargino rest frame, $E^*_{W^\pm}$ and $p^*_{W^\pm}$, can be
derived from the $\tilde{\chi}^\pm_1, \tilde{\chi}^0_1$ and $W^\pm$
masses. Furthermore, the electric charge of the individual $W^\pm$ and
$\tilde{\chi}^\pm_1$ can be identified by tagging the electric charge of
the $W^\mp$ from the other chargino $\tilde{\chi}_1^\mp$ decay through the
leptonic mode $W^\mp\to\ell^\mp\nu_\ell$.  All these features can be used to
study the $\tilde{\chi}^\pm_1$-polarization through the angular distribution in
the two-body decays $\tilde{\chi}^\pm_1\to W^\pm\tilde{\chi}^0_1$. For the
chargino as a spin-1/2 particle the decay distribution is linear in
$\cos\theta^*_{W^\pm}$:
\begin{eqnarray}
\frac{1}{d\Gamma}\frac{\Gamma}{d\cos\theta^*_{W^\pm}}
      \left[\tilde{\chi}^\pm_1\to W^\pm\tilde{\chi}^0_1\right]
= \frac{1}{2}\left(1 + \langle\kappa_{W^\pm}\rangle\cos\theta^*_{W^\pm} \right)
\label{eq:W_angle_distribution}
\end{eqnarray}
where the coefficient $\langle\kappa_{W^\pm}\rangle$ is the product
of the $\tilde{\chi}^\pm_1$ polarization averaged over the
production distribution and the $\tilde{\chi}^\pm_1$ polarization
analysis power of the decay mode
$\tilde{\chi}^\pm_1\to W^\pm\tilde{\chi}^0_1$; for details see
Ref.~\cite{Choi:1998ut}. The two coefficients $\langle \kappa_{W^\pm}\rangle$
are identical as a consequence of CP symmetry. The average helicities of
$\tilde{\chi}^-_1$ and $\tilde{\chi}^+_1$ have the same magnitude but opposite
sign in $e^+e^-$ annihilation for arbitrary beam polarization.
For non-zero $\langle\kappa_{W^\pm}\rangle$ the decay angular distributions
provide a unique signal for the spin $J=1/2$ of the chargino
$\tilde{\chi}^\pm_1$.\s

\begin{figure}[tbh!]
\begin{center}
\includegraphics[width=17.cm,height=15.5cm,angle=0]{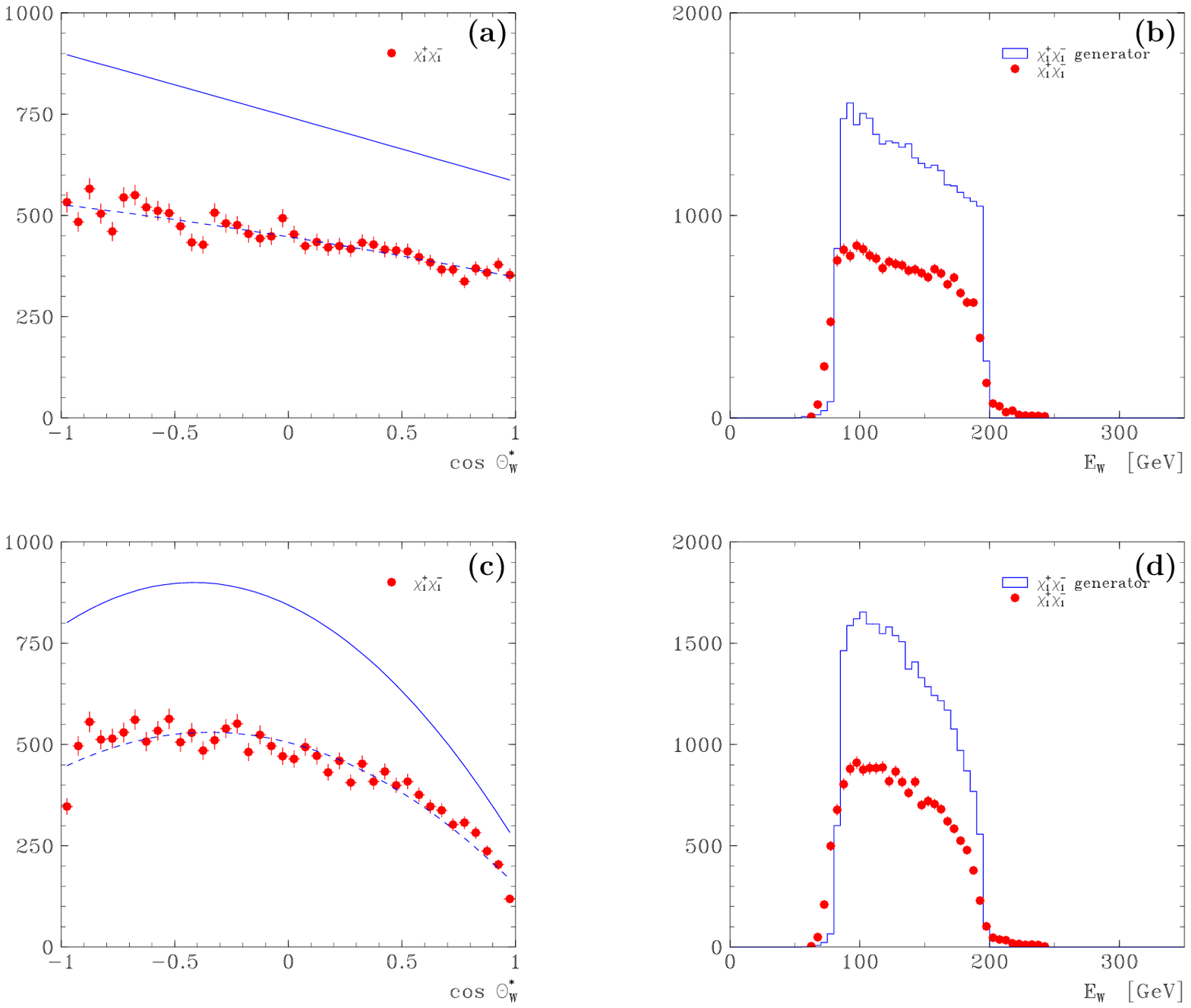}
\end{center}
\vskip -0.5cm
\caption{\it Chargino production $\ee\to\cx^+_1\cx^-_1\to W^+\nt_1\;W^-\nt_1$
  with $W^\pm$ decay angle distributions $\cos\theta^*_W$ in the $\cx^\pm_1$
  rest frame (left panels) and $W^\pm$  energy distributions $E_W$ in the
  laboratory system (right panels). The plots (a) and (b) do not include
  QED radiation effects while the plots (c) and (d) include initial state
  photon radiation and beamstrahlung. Spectra at generator level (full curves
  and histograms) are compared with events simulated and reconstructed
  in the detector (dots), the dashed curves indicate fits to the data.
  The simulation is based on polarized beams with $(\cP_{e^-},\cP_{e^+})
  = (-0.8, +0.6)$ at $\sqrt{s}=700\,\GeV$ and $\cL=500\,\fbi$.}
\label{fig:chargino_decay}
\end{figure}

The $W^\pm$ decay angular distribution is studied in an experimental simulation
of $\ee\to\cx^+_1\cx^-_1$ production at $\sqrt{s}=700\,\GeV$,
assuming the integrated luminosity of $\cL=500\,\fbi$.
With beam polarizations $(\cP_{e^-},\cP_{e^+}) = (-0.8, +0.6)$ the cross
section amounts to $\sigma = 150\,\fb$. The event signature is a reconstructed
hadronic decay $W\to q\bar q'$, a lepton from the decay $W\to \ell\nu$
($\ell = e, \mu, \tau$) to select clean events, and large missing energy of
$E^{\rm miss} > \sqrt{s}/2$. With a typical selection efficiency
$\epsilon\simeq 0.6$ and a combined branching ratio $\cB=0.44$ prolific event
rates are expected. Background from other SM or SUSY processes is estimated
to be small and will not be considered further. The chargino sample may
be tripled by including events where both $W'$s are allowed to decay to hadrons.
However, the background will also increase due to false combinations of jets
in reconstructing the two $W'$s and due to background from $\nt_2\nt_2$ production
(see above). Such a study goes beyond the aim of the present paper.\s

The basis of the analysis is Eq.$\,$(\ref{eq:energy_angle_relation}) which relates
the $W$ energy $E_W$ in the laboratory system with the decay angle
$\cos\theta^*_{W^\pm}$
in the $\cx^\pm_1$ rest frame.\s

Figures~\ref{fig:chargino_decay}(a) and (b) show the angular distribution and energy
spectrum for the hypothetical case that no QED radiation degrades the nominal
production energy. The linear $\cos\theta^*_{W^\pm}$ dependence is clearly seen at
generator level as well as after detector simulation. A fit of the data to the
function $d\sigma/d \cos\theta^*_{W^\pm} \sim 1 + a^0_1\cos\theta^*_{W^\pm}
+ a^0_2\cos2\theta^*_{W^\pm}$ yields $a^0_1 = -0.213\pm 0.010$ and $a^0_2 =
-0.001\pm 0.010$. This value is consistent with the theoretical expectation of
Eq.$\,$(\ref{eq:W_angle_distribution}) with $\langle \kappa_{W^\pm}\rangle = -0.216$
and demonstrates that distortions due to event selection criteria and detector
effects are small. The same tendency is
observed in the energy distribution Fig.$\,$\ref{fig:chargino_decay}(b)
which falls linearly with $E_W$, as compared with a flat distribution for
unpolarized charginos.\s

In the more realistic situation that initial state photon radiation (ISR) and
beamstrahlung decrease the $\cx^+_1\cx^-_1$ production energy, the angular
distribution is no longer linearly falling, as shown in
Fig.$\,$\ref{fig:chargino_decay}(c). Considerable depletions at
$\cos\theta^*_{W^\pm}\to \pm 1$ are observed since the constraint
$E_{\cx_1^\pm}$ is not always valid. However, since both ISR and beamstrahlung
effects can be calculated theoretically and measured precisely, they can be
unfolded from the data, e.g. by applying a bin-by-bin correction (like in the
present analysis) or a matrix inversion procedure. Fitting of the QED
corrected angular distribution (not shown) to the form
\begin{eqnarray}
\frac{d \sigma^{\rm exp}}{d \cos\theta^*_{W^\pm}}\, \sim\,
   1 + a_1\cos\theta^*_{W^\pm} + a_2\cos 2\theta^*_{W^\pm}
\end{eqnarray}
results in coefficients
\begin{eqnarray}
a_1 = -0.203\pm 0.020\quad \mbox{and}\quad a_2 = -0.001\pm 0.020
\end{eqnarray}
These values are consistent with the input parameters
and confirm with high precision the linear dependence on $\cos\theta^*_W$
characteristic for polarized spin 1/2 chargino production, while
higher spin-$J$ states would generate the angular distribution
\begin{eqnarray}
\frac{d\sigma}{d\cos\theta^*_{W^\pm}} \,\sim\, 1+\sum^{2J}_{n=1} a_n
 \cos n\,\theta^*_{W^\pm}
\end{eqnarray}
A sensitivity of a few percents to any term in addition to the linear term
can be reached which is an important bound in discriminating against
higher spin $J>1/2$ states. Similar distortions due to QED radiation
can be seen in the energy spectrum of
Fig.$\,$\ref{fig:chargino_decay}(d) which is shifted towards lower values and
is considerably depopulated at the maximum energy.\s

\noindent
{\bf (b)} \underline{Neutralino $\tilde{\chi}^0_2$}:\\
The distribution of the polar angle in neutralino production $e^+e^-\to
\tilde{\chi}^0_2\tilde{\chi}^0_2$ with two identical Majorana particles in the
final state is forward-backward symmetric. The $\tilde{\chi}^0_2$ polarization,
being non-zero for fixed polar angle, is asymmetric if the angle is varied from
the forward to the backward direction \cite{choi_neutralino}.
The polarization degree can be enhanced by using polarized
electrons/positrons beams. \s

The $\tilde{\chi}^0_2$ polarization can be determined \cite{Bartl:2004ut} in
the two-body decay $\tilde{\chi}^0_2\to Z\tilde{\chi}^0_1$ if the $Z$ polarization
is measured in the leptonic decays $Z\to\ell^+\ell^-$. The measurement can also
be performed for the hadronic decays $Z\to c\bar{c}$ and $b\bar{b}$ with
$c$ and $b$ flavor tagging.\footnote{A large number of events will be required
for the leptonic decays because of the small branching ratio ($\sim 0.07$ for
$e/\mu$) and the small analysis power ($\sim\! -0.15$) as a result of the almost
pure axial-vector $Z\ell\ell$ coupling. By contrast, the hadronic decays have
four times larger branching ratios and much larger analysis powers ($\sim\!-0.92$
and $-0.67$ for $b$ and $c$ quarks) than the
leptonic decays.} \s

(Alternatively, if kinematically accessible,) the two-body leptonic decay
$\tilde{\chi}^0_2\to \tilde{\ell}^\pm\ell^\mp$\, can provide a powerful instrument
for determining the $\tilde{\chi}^0_2$ spin. The $\tilde{\chi}^0_2$ momenta can be
reconstructed, event by event, in $\tilde{\chi}^0_2$ pair production for
sequential $\tilde{\chi}^0_2$ leptonic decays because the two unknown
$\tilde{\chi}^0_1$ momenta can be fixed by measuring the four visible lepton
momenta in the cascade decays $\tilde{\chi}^0_2\to\tilde{\ell}^\pm\ell^\mp\to
\ell^+\ell^- \tilde{\chi}^0_1$ and $\tilde{\chi}^0_2\to\tilde{\ell'}^\pm
{\ell'}^\mp \to {\ell'}^+{\ell'}^-\tilde{\chi}^0_1$. Furthermore, the slepton
mode is a perfect polarization analyzer of the decaying neutralino. Explicitly,
the angular distribution in the rest frame of the decaying spin-1/2
neutralino $\tilde{\chi}^0_2$ is given by [see Ref.~\cite{cds}]
\begin{eqnarray}
\frac{1}{\Gamma}\frac{\Gamma}{d\cos\theta^*_{\ell^\mp}}\left[\tilde{\chi}^0_2\to
\tilde{\ell}^\pm_R\ell^\mp\right] =\frac{1}{2} \left(1\pm {\cal P}_{\tilde{\chi}^0_2}
\cos\theta^*_{\ell^\mp}\right)
\end{eqnarray}
where ${\cal P}_{\tilde{\chi}^0_2}$ is the degree of longitudinal $\tilde{\chi}^0_2$
polarization and $\theta^*_{\ell^\mp}$ the angle of the $\ell^\mp$ momentum
in the $\tilde{\chi}^0_2$ rest frame with respect to the $\tilde{\chi}^0_2$
momentum direction. \s

Therefore, the decays $\tilde{\chi}^0_2\to Z\tilde{\chi}^0_1$ and/or
$\tilde{\chi}^0_2\to\tilde{\ell}\ell$ do provide a unique signal for the spin
$J=1/2$ of the neutralino $\tilde{\chi}^0_2$.\s

\begin{table}[thb]
\caption{\label{tab:summary} {\it The threshold behavior and the
angular distribution of SUSY and UED particle pair production, and
the general characteristics of spin-$J$ field theories.
All the characteristics refer to diagonal
pair production. B and F$_{D,M}$ generically denote bosons and Dirac,
Majorana fermions; $[s]$ $s$-channel exchange only, $[s,t,u]$ potentially
all three exchange mechanisms. The parameters $\kappa$ [$\kappa \neq -1$]
depend on mass ratios and particle velocities $\beta$; note that, especially,
$\kappa=1$ for Majorana fermions and for self-conjugate bosons [spin $\geq 1$]
in $[s]$-channels -- Notice the uniqueness of spin-0 assignments by measurements
of the polar angle distribution in the slepton sector. Neither threshold
excitation nor angular distributions are sufficient in the chargino/neutralino
sector and final state analyses must be performed to determine the spin-1/2
quantum numbers.}}
\mbox{ }\\[-0.5cm]
\begin{center}
\begin{tabular}{|c||c|cc|cc|}
\hline
\multicolumn{6}{|c|} {Threshold Excitation and Angular Distribution}\\
 \hline \hline
SUSY & particle   & $\tilde{\mu}$  & $\tilde{e}$ & $\tilde{\chi}^\pm$
     & $\tilde{\chi}^0$ \\
{ }  &  spin &   0      &   0   & 1/2
     & 1/2 \\
\cline{2-6}
{ }  & $\sigma_{thr}$ & $\beta^3$      & $\beta^3$   & $\beta$
     & $\beta^3$ \\
{ }  & $\theta$ dep.  & $\sin^2\theta$  &  thr: $\sin^2\theta$
     & thr: isotropic  & thr: $1+\kappa \cos^2\theta$ \\ \hline \hline
UED  & particle  & $\mu_1$ & $e_1$  & $W^\pm_1$ & $Z_1$ \\
{ }  & spin &  1/2    &  1/2   &  1        &   1 \\
\cline{2-6}
{ }  & $\sigma_{thr}$ & $\beta$      & $\beta$   & $\beta$
     & $\beta$ \\
{ }  & $\theta$ dep.  & $1+\kappa^2\cos^2\theta$  &  thr: isotropic
     & thr: isotropic  & thr: isotropic \\ \hline \hline
General  & particle  & $B[s]$ & $B[s,t,u]$  & $F_{D,M}[s]$ & $F_{D,M}[s,t,u]$ \\
{ }  & spin &  $\geq 1$    &  $\geq 1$   &  $\geq 1/2$  & $\geq 1/2$ \\
\cline{2-6}
{ }  & $\sigma_{thr}$ & $\beta^3$     & $\beta$   & $\beta,\beta^3$
     & $\beta,\beta^3$ \\
{ }  & $\theta$ dep.  & $1+\kappa\cos^2\theta$  &  thr: isotropic
     & $1+\kappa \cos^2\theta$  & thr: $1+\kappa \cos^2\theta$\\ \hline
\end{tabular}
\end{center}
\end{table}
%

\section{Summary}
\setcounter{equation}{0}

It is apparent from the preceding discussion that the model-independent
determination of the spin quantum numbers of supersymmetric particles
is a complex task, with the degree of complexity depending
on the nature of the particle. Threshold
excitation and angular distributions in pair production as well as
angular correlations in particle decays provide the signals for
experimental spin measurements. \s

The predictions for the threshold excitation and the angular distributions
in the production processes of supersymmetric particles are summarized in
Table~\ref{tab:summary}. They are confronted with predictions for particles
in models of universal extra space dimensions and with general analyses based on
the non-vanishing of the magnetic dipole moments of all spin $> 0$ particles.\s

Examining these results it turns out that the $\sin^2\theta$ law for the
production of \underline{spin-0 sleptons} [for selectrons close to threshold]
is a unique signal of the spin-0 character. While the observation of the
$\sin^2\theta$ angular distribution is sufficient for sleptons, the $\beta^3$
onset of the excitation curve is a necessary but not a sufficient condition for the
spin-0 character. Thus the spin determination in the slepton sector
is conceptually very simple at $e^+e^-$ colliders.  \s

This simple pattern in the slepton sector must be contrasted with the more
involved pattern in the spin-1/2 \underline{chargino/neutralino} sector.
Neither the onset of excitation curves nor the angular distributions in the
production processes provide unique signals of the spin quantum numbers.
However, decay angular distributions, $\sim |d^J(\theta^\ast)|^2$,
do provide a unique signal for the chargino/neutralino spin $J = 1/2$, albeit
at the expense of more involved experimental analyses. Using polarized
electron/positron beams will in general assure that the decaying spin-1/2
particle is polarized; reasonable polarization analysis power is guaranteed
in many decay processes.\s

{\it In toto}. The spin of sleptons and charginos/neutralinos can be
determined in a model-independent way at $e^+e^-$ colliders. Similar
methods as elaborated for sleptons can be applied in the squark sector
while gluinos will demand a methodologically separate analysis. \s  \vskip 12mm

\subsection*{Acknowledgments}

The work was supported in part by the Korea Research Foundation
Grant (KRF-2006-013-C00097), by KOSEF through CHEP at Kyungpook
National University, by the Deutsche Forschungsgemeinschaft
and by the Grant-in-Aid for Scientific Research (17540281 and
18340060) from MEXT, Japan.
S.Y.C. thanks for support during his visit to DESY, while P.M.Z.
gratefully acknowledges the warm  hospitality extended to him by KEK.

\vspace{1.cm}

\setcounter{equation}{0}
\setcounter{section}{1}
\def\thesection{\Alph{section}}
\renewcommand{\theequation}{{\rm \thesection.\arabic{equation}}}

\section*{Appendix: Cross sections with polarized beams}

Polarized electron and positron beams at $e^+e^-$ colliders are useful for
diagnosing the properties of supersymmetric particles and for
unraveling the underlying structure of the SUSY theory \cite{moortgat-pick}.
In this Appendix we present the general formulae for the production cross sections
of R-type smuon/electron pairs, and chargino and neutralino pairs
in $e^+e^-$ annihilation with polarized electron and positron beams.\s

For longitudinal $e^\pm$ beam polarizations ${\cal P}_{e^\pm}$ the polarized
production cross sections for R-type smuon and selectron pairs in $e^+e^-$
annihilation are given in terms of the charges $Q_R$ and $Q_L$ by
\begin{eqnarray}
 \frac{d\sigma}{d\cos\theta}
 \{\tilde{\ell}^+_R\tilde{\ell}^-_R\}
  \, =\, \frac{3}{32}\,\sigma_0\, \beta^3\sin^2\theta\,
     \bigg[(1-{\cal P}_{e^-}{\cal P}_{e^+})(|Q_R|^2+|Q_L|^2)
           +({\cal P}_{e^-}-{\cal P}_{e^+})(|Q_R|^2-|Q_L|^2)\bigg]
\end{eqnarray}
where $\tilde{\ell}_R=\tilde{\mu}_R, \tilde{e}_R$ and
$\sigma_0=4\pi\alpha^2/3s$ is the standard normalization cross section of $e^+e^-$
annihilation. The expressions of the generalized charges $Q_R$ and $Q_L$ for
R-type smuon-  and selectron-pair production can be found in
Eqs.$\,$(\ref{eq:smuon_Q_R}/\ref{eq:smuon_Q_L})
and (\ref{eq:qr_ser}/\ref{eq:ql_ser}),
respectively. For right/left polarized electrons ${\cal P}_{e^-}=\pm$ and unpolarized
positrons, the production cross sections
\begin{eqnarray}
&& \frac{d\sigma_R}{d\cos\theta}
 \{\tilde{\ell}^+_R\tilde{\ell}^-_R\}
  \, =\, \frac{3}{16}\,\sigma_0\, \beta^3\sin^2\theta\,|Q_R|^2 \\
&& \frac{d\sigma_L}{d\cos\theta}
 \{\tilde{\ell}^+_R\tilde{\ell}^-_R\}
  \, =\, \frac{3}{16}\,\sigma_0\, \beta^3\sin^2\theta\,|Q_L|^2
\end{eqnarray}
project out the bilinear $R$ and $L$ charges $Q_R$ and $Q_L$. \s

The production cross sections for chargino- and neutralino-pairs
in $e^+e^-$ annihilation with polarized electron and positron beams are
given by
\begin{eqnarray}
\mbox{ }\hskip -0.8cm \frac{d\sigma}{d\cos\theta}
\left\{\tilde{\chi}_i\tilde{\chi}_j\right\}
 \!\!\!\!&=&\!\!\! \frac{3}{8}\sigma_0\,\beta\,
     \bigg[(1-{\cal P}_{e^-}{\cal P}_{e^+})
           \left\{[1-(\mu^2_i-\mu^2_j)^2+\beta^2\cos^2\theta] Q_1
       +4\mu_i\mu_j Q_2
       +2\beta\cos\theta Q_3\right\} \nonumber\\
 &&{ }\hskip 0.6cm +({\cal P}_{e^-}-{\cal P}_{e^+})
       \left\{[1-(\mu^2_i-\mu^2_j)^2+\beta^2\cos^2\theta] Q'_1
       +4\mu_i\mu_j Q'_2
       +2\beta\cos\theta Q'_3\right\}\!\bigg]
\end{eqnarray}
where the $P$-even and $P$-odd quartic charges  $Q_i$ and $Q'_i$
($i=1,2,3$) are defined in terms of the bilinear charges $Q_{\alpha\beta}$
($\alpha,\beta=L,R$) as
\begin{eqnarray}
&& Q^{(\prime)}_1 \,=\, \frac{1}{4}\left[|Q_{RR}|^2+|Q_{RL}|^2
                           \pm |Q_{LR}|^2\pm |Q_{LL}|^2\right] \nonumber\\
&& Q^{(\prime)}_2 \,=\, \frac{1}{2}\real\left[Q_{RR}Q^*_{RL}
                                \pm Q_{LL}Q^*_{LR}\right] \nonumber\\
&& Q^{(\prime)}_3 \,=\, \frac{1}{4}\left[|Q_{RR}|^2-|Q_{RL}|^2
                           \pm |Q_{LR}|^2\mp |Q_{LL}|^2\right]
\label{eq:quartic_charges}
\end{eqnarray}
The explicit form  of the bilinear charges $Q_{\alpha\beta}$ for the production
of the chargino pair $\tilde{\chi}^+_1\tilde{\chi}^-_1$ and the neutralino pairs
$\tilde{\chi}^0_i\tilde{\chi}^0_j$ is given in
Eqs.$\,$(\ref{eq:chargino_bilinear_charges}) and
(\ref{eq:neutralino_bilinear_charges}), respectively. Polarized
electrons combined with unpolarized positrons,
\begin{eqnarray}
\frac{d\sigma_R}{d\cos\theta}
\left\{\tilde{\chi}_i\tilde{\chi}_j\right\}  \!&=&\!
\frac{3}{16}\sigma_0\,\beta\,
     \bigg[\, [1-(\mu^2_i-\mu^2_j)^2+\beta^2\cos^2\theta]\, (|Q_{RR}|^2+|Q_{RL}|^2)
        \nonumber\\
     &&{ }\hskip 1.5cm  +8\mu_i\mu_j\, \real(Q_{RR}Q^*_{RL})
       +2\beta\cos\theta (|Q_{RR}|^2-|Q_{RL}|^2) \bigg]\\
\frac{d\sigma_L}{d\cos\theta}
\left\{\tilde{\chi}_i\tilde{\chi}_j\right\}
    \!&=&\! \frac{3}{16}\sigma_0\,\beta\,
     \bigg[\, [1-(\mu^2_i-\mu^2_j)^2+\beta^2\cos^2\theta]\, (|Q_{LR}|^2+|Q_{LL}|^2)
       \nonumber\\
      &&{ }\hskip 1.5cm  +8\mu_i\mu_j\, \real(Q_{LR}Q^*_{LL})
       +2\beta\cos\theta (|Q_{LR}|^2-|Q_{LL}|^2) \bigg]
\end{eqnarray}
project out the bilinear charges $Q_{Rk}$ and $Q_{Lk}$ for the $\tilde{\chi}$
chirality $k=R,L$. \s

\vskip15mm


\end{document}